\documentclass[sigconf]{acmart}

\AtBeginDocument{%
  \providecommand\BibTeX{{%
    \normalfont B\kern-0.5em{\scshape i\kern-0.25em b}\kern-0.8em\TeX}}}

\copyrightyear{2021}
\acmYear{2021}
\setcopyright{acmcopyright}\acmConference[ARES 2021]{The 16th International Conference on Availability, Reliability and Security}{August 17--20, 2021}{Vienna, Austria}
\acmBooktitle{The 16th International Conference on Availability, Reliability and Security (ARES 2021), August 17--20, 2021, Vienna, Austria}
\acmPrice{15.00}
\acmDOI{10.1145/3465481.3470102}
\acmISBN{978-1-4503-9051-4/21/08}

\usepackage{csquotes}
\usepackage{caption}
\usepackage{subcaption}

\begin{document}

\title{Modeling of Personalized Privacy Disclosure Behavior: A Formal Method Approach}

\author{A K M Nuhil Mehdy}
\affiliation{%
  \institution{Computer Science Department, Boise State University}
  \streetaddress{1910 University Dr}
  \city{Boise}
  \state{Idaho}
  \country{USA}
  \postcode{83706}
}
\email{akmnuhilmehdy@u.boisestate.edu}

\author{Hoda Mehrpouyan}
\affiliation{%
  \institution{Computer Science Department, Boise State University}
  \streetaddress{1910 University Dr}
  \city{Boise}
  \state{Idaho}
  \country{USA}}
\email{hodamehrpouyan@boisestate.edu}



\begin{abstract}
In order to create user-centric and personalized privacy management tools, the underlying models must account for individual users' privacy expectations, preferences, and their ability to control their information sharing activities. Existing studies of users' privacy behavior modeling attempt to frame the problem from a request's perspective, which lack the crucial involvement of the information owner, resulting in limited or no control of policy management. Moreover, very few of them take into the consideration the aspect of correctness, explainability, usability, and acceptance of the methodologies for each user of the system. In this paper, we present a methodology to formally model, validate, and verify personalized privacy disclosure behavior based on the analysis of the user's situational decision-making process. We use a model checking tool named UPPAAL to represent users' self-reported privacy disclosure behavior by an extended form of finite state automata (FSA), and perform reachability analysis for the verification of privacy properties through computation tree logic (CTL) formulas. We also describe the practical use cases of the methodology depicting the potential of formal technique towards the design and development of user-centric behavioral modeling. This paper, through extensive amounts of experimental outcomes, contributes several insights to the area of formal methods and user-tailored privacy behavior modeling.

\end{abstract}




\maketitle

\section{Introduction}
\label{section:introduction}

Privacy in the information domain refers to the right of a person to monitor and control the processing, exposition, and preservation of information about themselves. \cite{lu2014verification}. Accordingly, the responsibility is on the user themselves to take control of what kind of information should be shared with whom, when, and how \cite{westin1968privacy, nuhil2019privacy, joshaghani2019formal,mehrpouyan2017measuring}. However, for an individual, it is quite cumbersome and difficult to manage and control their information sharing preferences \cite{xiao2006personalized}. This is because different devices, applications, and software require different privacy configurations from the users, and most importantly they are not designed to be personalized or assisting. Therefore, it is important than ever before to develop and provide suitable tools and algorithms to the users so that they can define, manage, and make the best use of their privacy preferences with ease. Existing methodologies and protocols intend to tackle this problem by employing technique such as access control policies \cite{osborn2000configuring, shen2006attribute}, machine readable privacy policy languages \cite{cranor2002web, ashley2003enterprise}, formal methods \cite{aucher2011dynamic,breaux2014eddy}, machine learning \cite{costante2012machine,tesfay2018read,mehdy2020user}, etc.  However, most of the works attempt to frame the problem from a request's perspective which lack the crucial involvement of the information owner, resulting in limited or no control of policy adjustment. Moreover, a very few of them take into consideration the aspect of personalization and explainability of such tools. Most importantly, while there is a significant amount of research aimed at design and development of privacy management tools and techniques, 'their practical usability and acceptance remains an important challenge' \cite{kurkovsky2007classification}.

Therefore, this paper applies model-based analysis to personalize privacy behavior which answers two key research question: how to model privacy behaviour and how to use this privacy behavior model for analysis. We decomposed this problem into three subcategories: (I) Identification of relevant privacy behavior and situational factors, (II) applying proper modeling techniques, (III) validating the models.

As part of model-based approach, we focuse on formal methods that are concerned with modeling, specifying, and verifying any systems using mathematical techniques otherwise known as model checking \cite{clarke1996formal}. A system could be physical or conceptual comprised of interconnected components such as processes, states, nodes, etc. Model checking is an automated approach to verify that a model of a system, usually a finite-state machine, satisfies a set of desired properties (i.e., requirement specifications) written in a temporal logic \cite{grumberg1999model}. This is achieved by exhaustively searching a system's state space in order to determine if these criteria hold. If there is a violation, an error trace is produced (i.e., a counterexample). Model checkers take system description (i.e., formal model) and a set of requirements as input and reason whither the requirements are satisfied or not. In privacy literature, human decision making, in other words, individual's intention to disclose private information is also considered as process which involves different components, otherwise known as influential factors \cite{ajzen1991theory}. When the number of factors is large, doing manual specification and testing of the privacy policies is difficult. It is also possible that subtle conditions get unnoticed. Again, a way to tackle this problem to a certain extent, is the use of mathematically-based techniques. Hence, we adopt the analogy of finite state machines from the theory of computation and aim to model human privacy disclosure behavior based on this specific formalism technique. 

That being said, to learn user's privacy behavior towards the development of user-specific models, it is important to investigate the factors and parameters that influence users to make dynamic privacy decisions \cite{ajzen1991theory, petronio2015communication, laufer1977privacy, lwin2003model}. The decision to exchange private information, as well as the risk perceptions that drive this decision, differs from situation to situation. Various considerations, such as the type of information, the receiver of the information, and the source of confidence underlying the reason for sharing, all play a role in the decision making process \cite{john2011strangers, simonson1992choice}. Moreover, risk assessment, potential risks consideration, and alternate exploration are all part of the process of deciding what to do in a specific situation \cite{west2009weakest, afifi2000motivations}. Additionally, individual variations in demographics, personality traits, and decision-making styles as well as their effect on users' privacy-related habits must be studied before developing any behavioral model. Therefore, we work on a dataset from \cite{mehdy2021privacy} which was obtained by conducting a custom designed survey on Amazon Mechanical Turk \footnote{A crowd sourcing website for businesses and researchers to hire remotely located "crowdworkers" to perform on-demand tasks such as survey, data labeling, etc.} (N=401) based on the theory of planned behavior (TPB) to measure the way users' perceptions of privacy factors and intent to disclose information are affected by three situational factors embodied hypothetical scenarios: information type, recipients' role, and trust source.

In this work, we chose to focus on user's situational decision-making process and represent our approach to formally model, validate, and verify personalized privacy behavior. We represent a scaled down version of our proposed methodology where we model each individual's privacy disclosure behavior where their disclosure decision merely rely on three factors--- information type, recipients' role, and trust source. Even though human decisions depend on many more factors, we chose this level of abstraction because the dataset in hand capture the users' privacy behavior based on these three factors. On the other hand, we wanted to evaluate our approach on top of ground truth dataset. Nevertheless, the methodology presented in this paper depicts potential of formalism towards the development of privacy management tools. This paper is the first to our knowledge to leverage extended version of automata based transitioned systems towards modeling individual's privacy behavior. This work provides insight into:

\begin{itemize}
\item Model-based analysis of personalized privacy behavior
\item Formulate personalized privacy policies
\item Detect and reason about unwanted disclosure behavior
\item Validate the proposed model-based approach and demonstrate its practicality
\end{itemize}

\section{Learning Privacy Preference}
\label{section:case-study}
In this work, we represent and evaluate our formal method approach to model users' privacy disclosure behavior based on a dataset that we obtained through a survey. We captured users' situational privacy decisions, through a custom scenario-based survey with 401 participants, each responding to a subset of 48 total unique scenarios. Every data point is referred to the responses to a series of questionnaires that assess participants' attitudes toward each situation, as well as their expectations of and willingness to reveal personal information in the given situation. By manipulating three situational factors: information type, recipient's role, and trust source, we use path analysis to model participants' privacy perceptions and plans, taking into account their assessments on subjective norm, perceived behavioral control, and attitude. This choice of factors is partly inspired from the theory of contextual integrity (CI) \cite{nissenbaum2004privacy, barth2006privacy}. The findings show how users make privacy decisions in a variety of contexts, as well as how situational factors influence users' views of privacy factors and their willingness to share private information. Most importantly, the results also reveal how every individual has their own preferences and concerns about disclosing their private information in certain situations. Therefore, this dataset best suit our personalized behavioral modeling experiment. The following sections describe the survey strategy and the data set in more detail.  

\subsection{Survey}
\label{section:survey}
After agreeing to participate in the survey, a person is given a series of eight hypothetical scenarios and asked to answer to them one by one. Each scenario places the subject in a position where he or she must choose whether or not to reveal the information embodied in that scenario. This includes the situational factors on which participants can place a high degree of confidence in their interpretation and decision on whether or not to disclose. We manipulate three situational factors to see how they affect participant responses:

\begin{description}
\item[Information Type (IT)]
    The type of the information that is illustrated in the scenario. Each scenario is about one of three information types: health, finance, or relationship.

\item[Recipient’s Role (RR)]    
    The type of the recipient, based on the relationship to the survey participant, to whom the information may be disclosed. We take into account four such recipient roles: family, friend, colleague, and online service (e.g., facebook, twitter, discussion forum, etc).

\item[Trust Source (TS)]
    From whom the participant got the motivation of disclosing the information to the recipient. We consider four trust sources: family, friend, expert (e.g., physician, counselor, financial adviser, etc), and self (i.e., searching the internet).
\end{description}

Different combinations of these factors yield a total of 48 (3*4*4) unique scenarios. For each of the combination, we prepare a scenario where a trust source encourages the participant to share the information with a recipient. We made every scenario as similar as possible to minimize extraneous variability while incorporating the factors in a natural and coherent manner in the hypothetical scenario. In other words, we made sure the framing of the scenarios does not become significantly different from each other so that only the factors get changed, and a proper parametric analysis is justified. An example scenario with \textit{health} as information type, \textit{friend} as trust source, and \textit{family member} as recipient's role could be:
\begin{displayquote}
Your doctor called and told you that your lab results came back positive for a disease. One of your friends suggested discussing the situation with a family member and asking their support, saying it could be helpful.
\end{displayquote}
Another unique scenario could be generated by changing the trust source from friend to family and recipient's role from family to online: 
\begin{displayquote}
Your doctor called and told you that your lab results came back positive for a disease. A family member suggested asking other patients and doctors on an online discussion forum, saying they have found it helpful for dealing with their similar condition.
\end{displayquote}

Every participant is assigned a set of 8 random scenarios with associated questionnaires. A participant has to read a given scenario and respond to all of the corresponding questions before proceeding to the next assigned scenario. We used rejection sampling to ensure that each user's 8 scenarios covered all 11 distinct factor levels at least once, ensuring a minimal degree of heterogeneity between their circumstances and, as a result, responses. To minimize order effects, we also randomly order the set of 8 scenarios for each participant. In the end, the individual completes a brief survey in which we intend to capture their general privacy attitudes regardless of any specific situation. This move is intended to capture expectations that are believed to be constant over time and do not alter in response to changing circumstances. Participants are asked to optionally enter their ethnicity, age group, country of origin, and period of residence in that country in the final phase of the survey for accumulating demographic information.

There are two sets of questions in the survey: i) scenario-specific questions (12 total) and ii) general attitude questions (4 total). For each of the eight scenarios allocated to each person, the first set of 12 questions is repeated. At the end of the survey, the second set of questions is presented. The scenario-specific questionnaire is inspired by \citet{heirman2013predicting}, and the second set of questions is inspired by prominent privacy research \cite{buchanan2007development, ackerman1999privacy}. Appendix \ref{appendix:survey-ui} shows a screenshot of the survey system representing 1 of 8 random scenarios given to a participant, and appendix \ref{appendix:survey-ui-ga} shows the screenshot representing the general attitude questions given to a participant at the end of the survey. Before the main survey, we conducted a pilot test with six of our research lab's colleagues. Their feedback was instrumental in resolving problems with the survey interface, user experience, and clarity of the scenarios and questionnaires. Later we used Amazon Mechanical Turk, an online crowd-sourcing marketplace, to find participants for the final survey. We looked for workers from the United States who are at least 18 years old with at least 95\% HIT (Human Intelligence Task) acceptance rate\footnote{whose previous works got approved by 95\% of the requesters.} and 50 hits approved.

\subsection{Dataset}
\label{section:dataset}
We employed a number of filters to ensure the quality of the data. First, we capture the time a participant spent on each scenario step and removed the data points from our analysis if the spent time was too low. Second, we randomly placed attention search questions in between survey questions. We also restricted repeated submissions from same participant by setting a browser cookie for 3 days after a satisfactory submission. The answers to the questions were translated into a numeric format (1 to 5) from the 5-point scale (ranging from Strongly Disagree to Strongly Agree). For the final decision question, we represent the Share and Not-Share options in logical numeric form, 1 and 0. In the end, we get 3208 data points, grouped by 401 participants, containing their information disclosure decisions based on different situational factors.

\subsection{Path Model for Privacy Behavior Analysis}
\label{section:path-model-pba}
\begin{figure}[tp]
    \centering
    \includegraphics[width=1\linewidth]{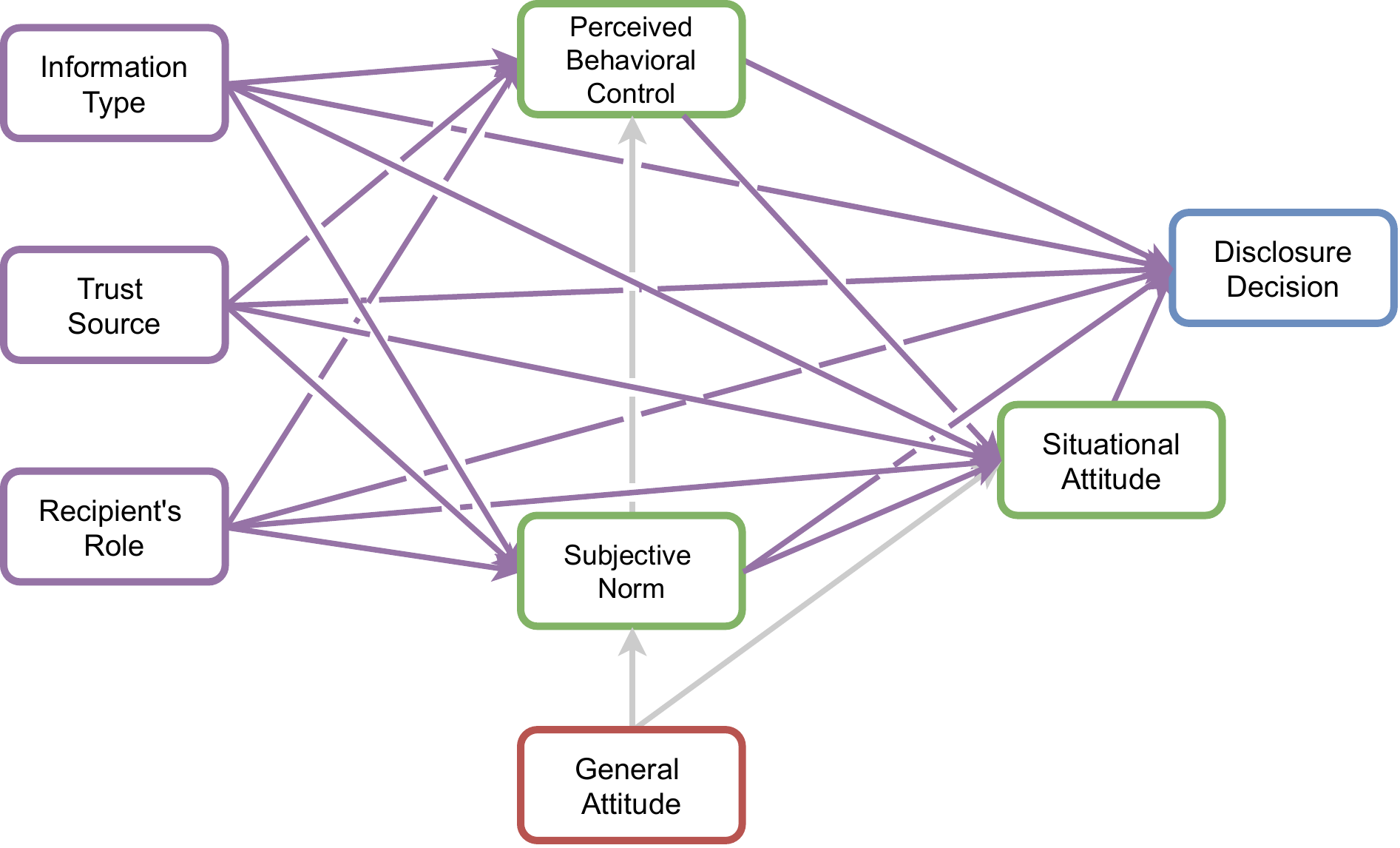}
    \caption{The Path Model for Analyzing Users' Privacy Decision-making Process.}
    \Description{Description}
    \label{figure:path-model}
    \vspace{-15pt}
\end{figure}

In one of our earlier works \cite{mehdy2021privacy}, we leveraged the data to measure users' behavioral intention and their situational perception of three constructs: attitude, subjective norm, and perceived behavioral control. These constructs and the path model is inspired from the theory of planned behavior (TPB) \cite{ajzen1991theory}. We also, incorporated the scenario factors--- information type, recipient's role, and trust source in our path analysis to measure the correlation of these factors with the information disclosure decision of the user. Figure \ref{figure:path-model} depicts the path model. The analysis results show that the path model fits the data very well with $\chi^2_{11}=12.017$, $p=0.3623$, $CFI=1.0$, $TLI=0.99$, $SRMR=0.008$, $RMSEA=0.005$, $90\%$  $CI=0.000$ to $0.020$. Also, the comparative fit index (CFI) and Tucker-Lewis index (TLI) values which ranges from 0 to 1 show near-perfect scores. 

Among all the path analysis results published in our work, one of the findings shows that there exist significant (indirect) effects of the scenario factors on the users' disclosure decisions. In Figure \ref{figure:path-model}, they refer to the paths from the purple leftmost boxes to the blue rightmost box via the mediator green boxes in between. These total effects describe \textit{how} users' intention changes from one scenario to another; the mediating TBP factors provide an explanation for \textit{why}. A few important findings include but not limited to--- with regard to the recipient's role in the scenario, compared to the recipient ``online service'', the odds of disclosure were estimated to be 16.6\% higher when the recipient was a family member and 12.9\% higher when the recipient was a friend; with regard to the type of information, compared to relationship information, the odds of disclosure were estimated to be 3.1\% lower when the scenario involved financial information and recipient was a family member and 5.1\% higher when the scenario involved health information, etc. These results indeed proof the influence of the situational factors towards users' disclosure decision and therefore act as the basic components of our formal privacy behavioral model.

\section{Formal Modeling}
\label{section:methodology}
This section describes the approach of developing the formal model of an user's privacy disclosure behavior by taking into account the privacy decisions made by that user. Our approach aims to address the issue of formally modeling the privacy behavior of an user which could be eventually utilized to develop a personalized privacy management system. The whole approach is divided into four main stages: i) observing user's historical sharing activity, ii) modeling users' personalized privacy behavior, iii) validating the model, iv) verifying the model given the privacy properties of the user. We have already detailed about the survey and the dataset in the earlier sections which refer to the first stage.

\subsection{Model Assumptions}
\label{section:model-assumption}
In this work, one of the main assumptions of the users' disclosure behavior is that user decides to share/not-share a specific type of information with a certain type of recipient(s) after being advised by a specific trust source. We represent these knowledge and the decision made by the user in the form of state model. Transitions between states occur with respect to a specific information type, trust source, and recipient's role. We also assume that there is no other factors/components involved in the user's decision making process. Additionally, we assume that user's behavior could conceivably be modeled as a finite state machine. This research utilizes finite state automata (FSA) extended with data variables to model the privacy disclosure behavior of the users.

\subsection{Model Paradigm}
\label{section:model-paradigm}
FSA as a chosen formalism allows for a design and development of a well-structured tools to conduct an automated analysis during the early stages of studying user's privacy behaviour. Accordingly, there are various tools for designing and verifying such FSA based formal models, i.e. NuSMV, PVS, Z3, and UPPAAL are a few of examples. We choose UPPAAL because of its ability to support model checking over network of automata using temporal logic \cite{behrmann2006tutorial}. UPPAAL also supports formalism through parallel compositionality among the automata. This modeling paradigm helps us to retrieve the traces of the transition while checking for a given query. Therefore, this modeling paradigm enables us to execute the requirements as temporal logic queries which in turns exhaustively check the satisfaction of the privacy properties.On the other hand, counter examples are provided to reason about privacy properties that are violated.

\section{Modeling in UPPAAL}
\label{section:modeling-uppaal}
The reasons for selecting UPPAAL is because UPPAAL provides a better graphical user-interface that allows for the development, modification, validation, and verification of any system model with drag and drop interface\cite{larsen1997uppaal}. In UPPAAL, a system is made up of several concurrent processes, each of which is modeled as an automaton. Each automaton has a set of locations otherwise known as states. Transitions between these states could be managed by guard and synchronization. A guard imposes  conditions on variables and clocks ensuring when the transition is enabled. Synchronization in UPPAAL enables two or more processes to communicate with each other based on a hand-shaking synchronization. Two actions are possible while a transition happens--- assignment of variables or reset of clocks. UPPAAL further extends timed automata with other types of data variables such as integer and Boolean towards developing a modeling language which is as close as a high level programming language \cite{larsen1997uppaal}. 

\begin{table*}[!ht]
  \caption{Disclosure Decisions by the User 89 Captured by the Survey}
  \label{tab:user-89-activities}
  \begin{tabular}{llllll}
    \toprule
    No&Scenario&IT&TS&RR&Share?\\
    \midrule
    1 & \begin{tabular}[c]{@{}l@{}}You recently had a very bad argument with your partner.\\ Your counsellor suggested sharing and discussing this matter\\ with a family member, saying they could support you.\end{tabular} & Relationship & Expert & Family & Yes\\
    \midrule
    2 & \begin{tabular}[c]{@{}l@{}}Your doctor called and told you that your lab results came back\\ positive for a disease. A family member suggested discussing\\ the situation with a family member and asking their support,\\ saying it could be helpful.\end{tabular} & Health & Family & Family & Yes\\
    \midrule
    3 & \begin{tabular}[c]{@{}l@{}}Your doctor called and told you that your lab results came back\\ positive for a disease. A family member suggested asking other\\ patients and doctors on an online discussion forum, saying they\\ have found it helpful for dealing with their similar condition.\end{tabular} & Health & Family & Online & No\\
    \midrule
    4 & \begin{tabular}[c]{@{}l@{}}You recently had a very bad argument with your partner. One\\ of your friends suggested asking on an online discussion forum\\ they use to get support from others, saying they have found it\\ helpful for dealing with their situation.\end{tabular} & Relationship & Friend & Online & No\\
    \midrule
    5 & \begin{tabular}[c]{@{}l@{}}Your doctor called and told you that your lab results came back\\ positive for a disease. You did some research and found that\\ people often find it helpful to get support from a colleague.\end{tabular} & Health & Self & Colleague & No\\
    \midrule
    6 & \begin{tabular}[c]{@{}l@{}}You received a notice from a collection agency saying you have\\ a debt which needs immediate attention. A family member\\ suggested asking on an online discussion forum they use to get\\ support from others, saying they have found it helpful for\\ managing a similar situation.\end{tabular} & Finance & Family & Online & No\\
    \midrule
    7 & \begin{tabular}[c]{@{}l@{}}You received a notice from a collection agency saying you have\\ a debt which needs immediate attention. Your financial advisor\\ suggested discussing the situation with a friend and asking\\ their support, saying it could be helpful.\end{tabular} & Finance & Expert & Friend & Yes\\
    \midrule    
    8 & \begin{tabular}[c]{@{}l@{}}You recently had a very bad argument with your partner. A family\\ member suggested sharing and discussing this matter with a\\ colleague, saying they could support you.\end{tabular} & Relationship & Family & Colleague & No\\
  \bottomrule
\end{tabular}
\end{table*}

\begin{figure}
    \centering
    \includegraphics[width=1\linewidth]{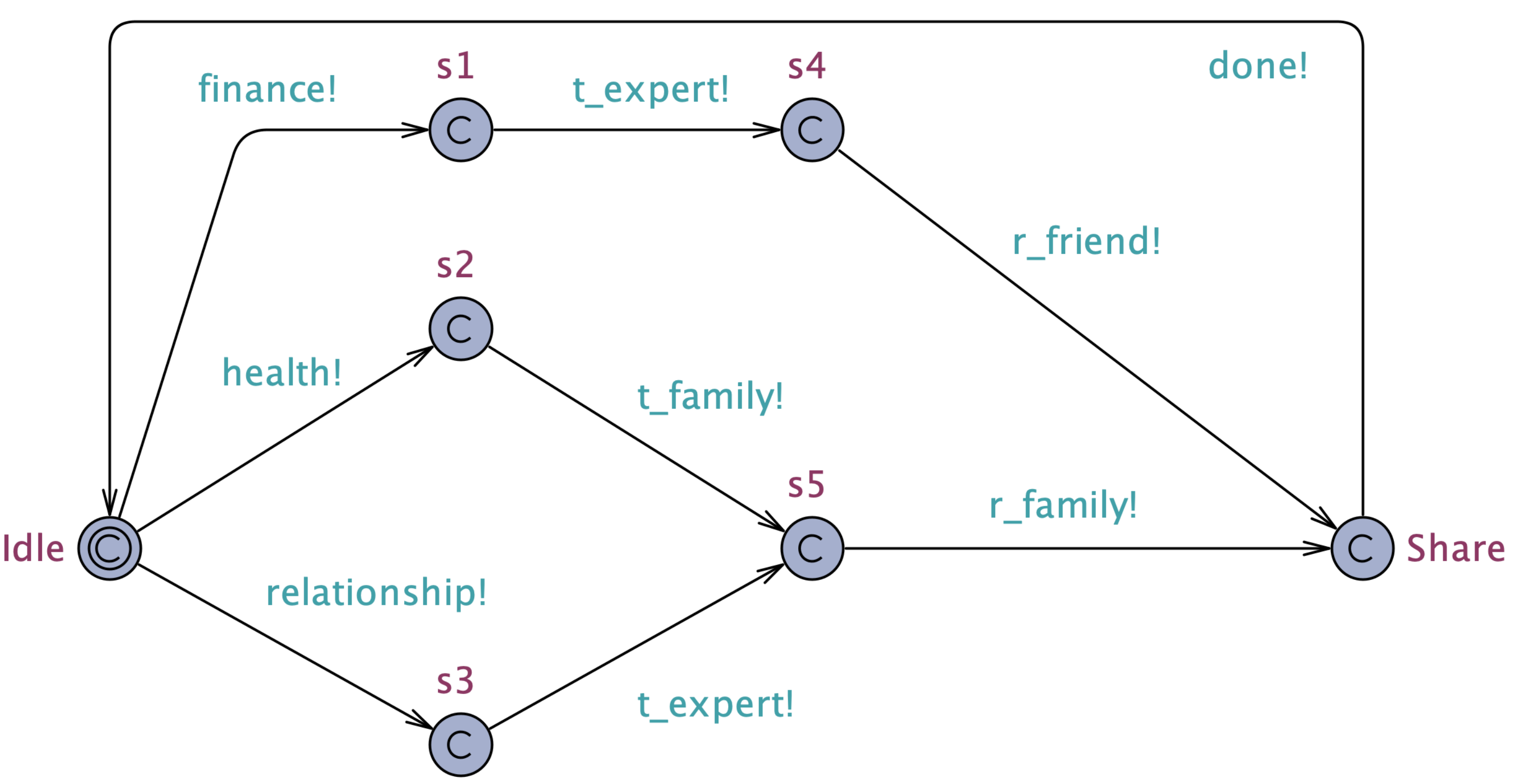}
    \caption{The Behavioral Model of User 89 Created in UPPAAL}
    \Description{Description}
    \label{figure:user-1-fsm}
    \vspace{-15pt}
\end{figure}

\begin{figure*}
     \centering
     \begin{subfigure}[b]{0.33\textwidth}
         \includegraphics[width=\textwidth]{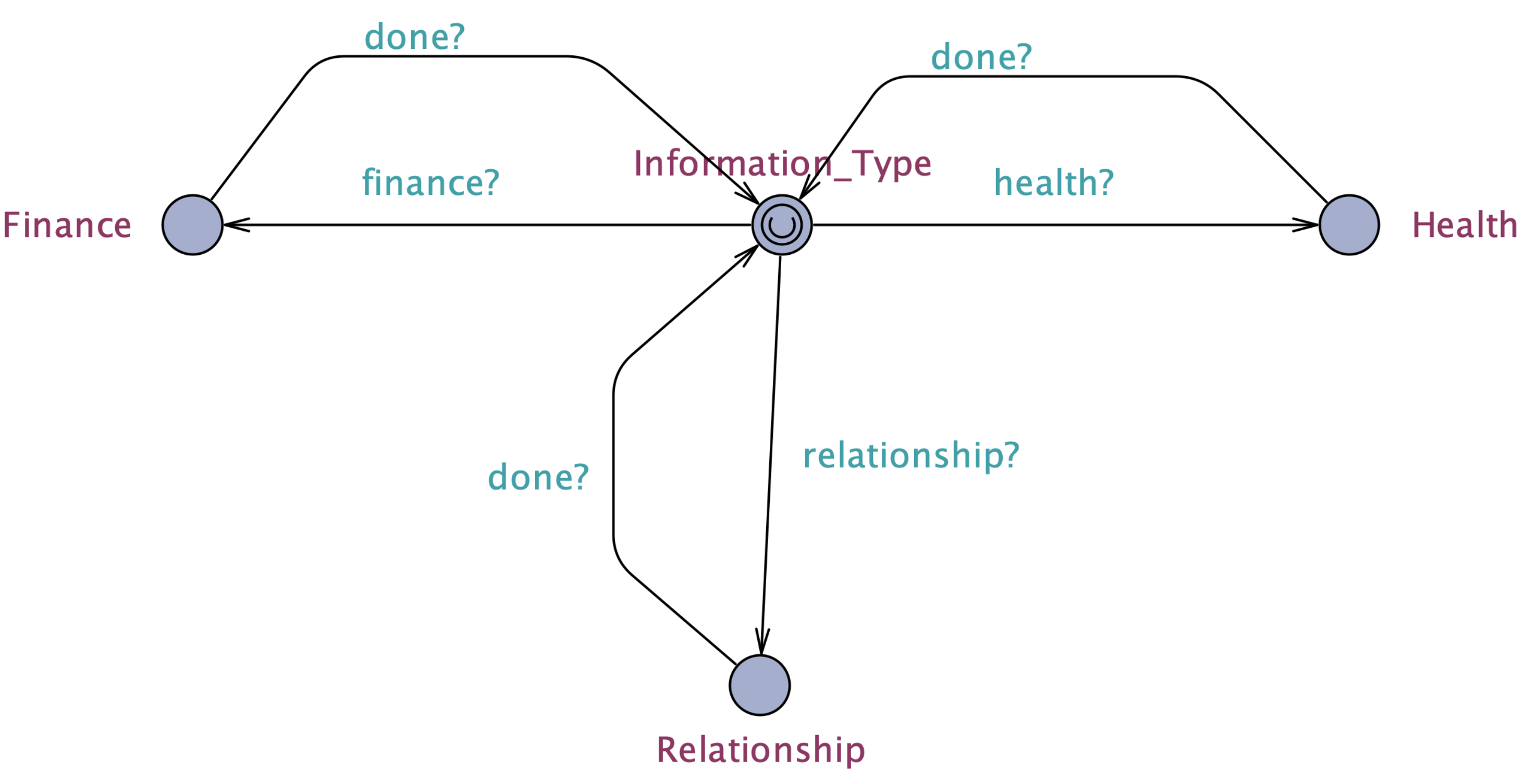}
         \caption{Information Type Observer}
         \label{fig:obs-it}
     \end{subfigure}
     \begin{subfigure}[b]{0.33\textwidth}
         \includegraphics[width=\textwidth]{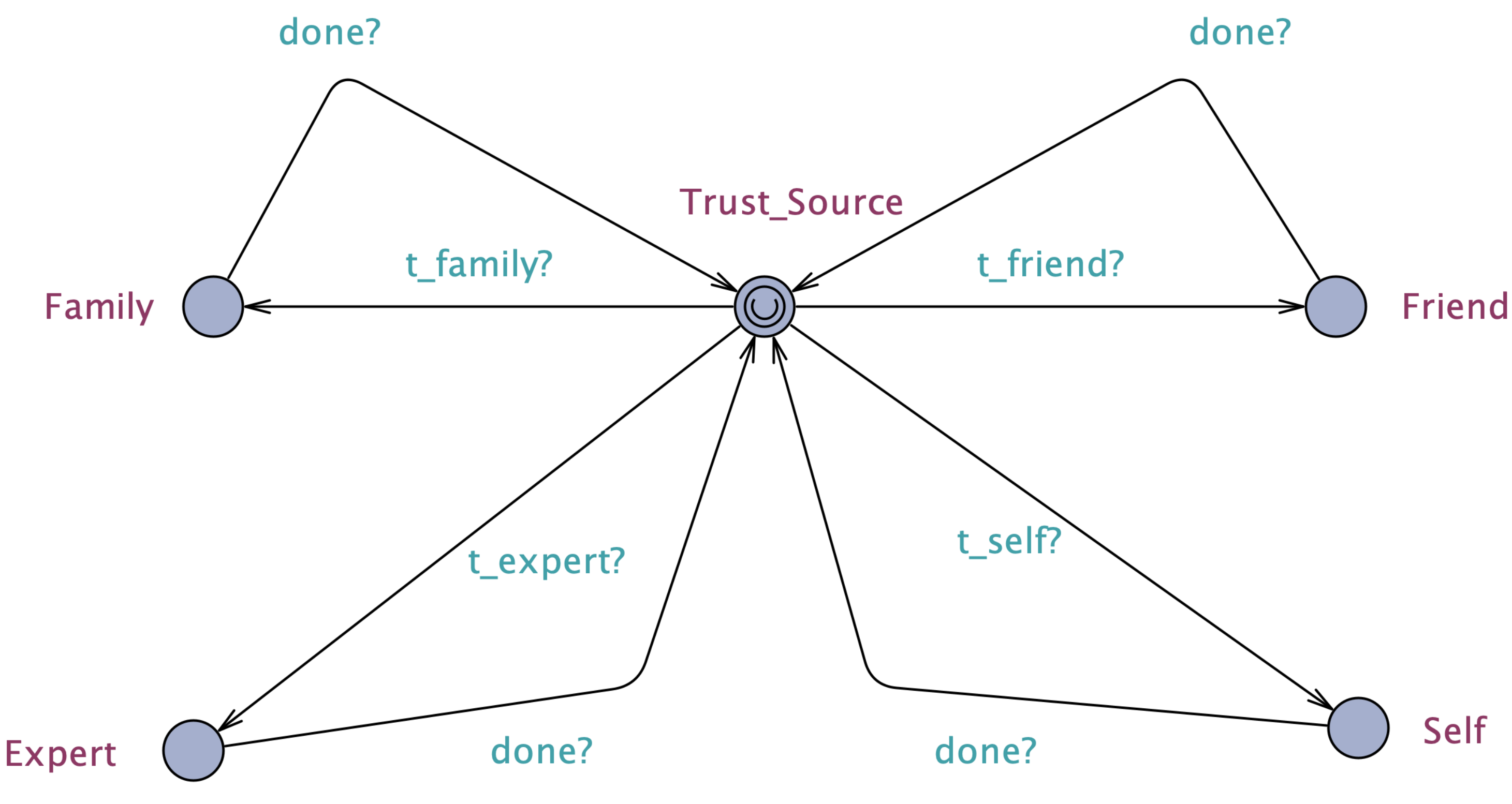}
         \caption{Trust Source Observer}
         \label{fig:obs-ts}
     \end{subfigure}
     \begin{subfigure}[b]{0.33\textwidth}
         \includegraphics[width=\textwidth]{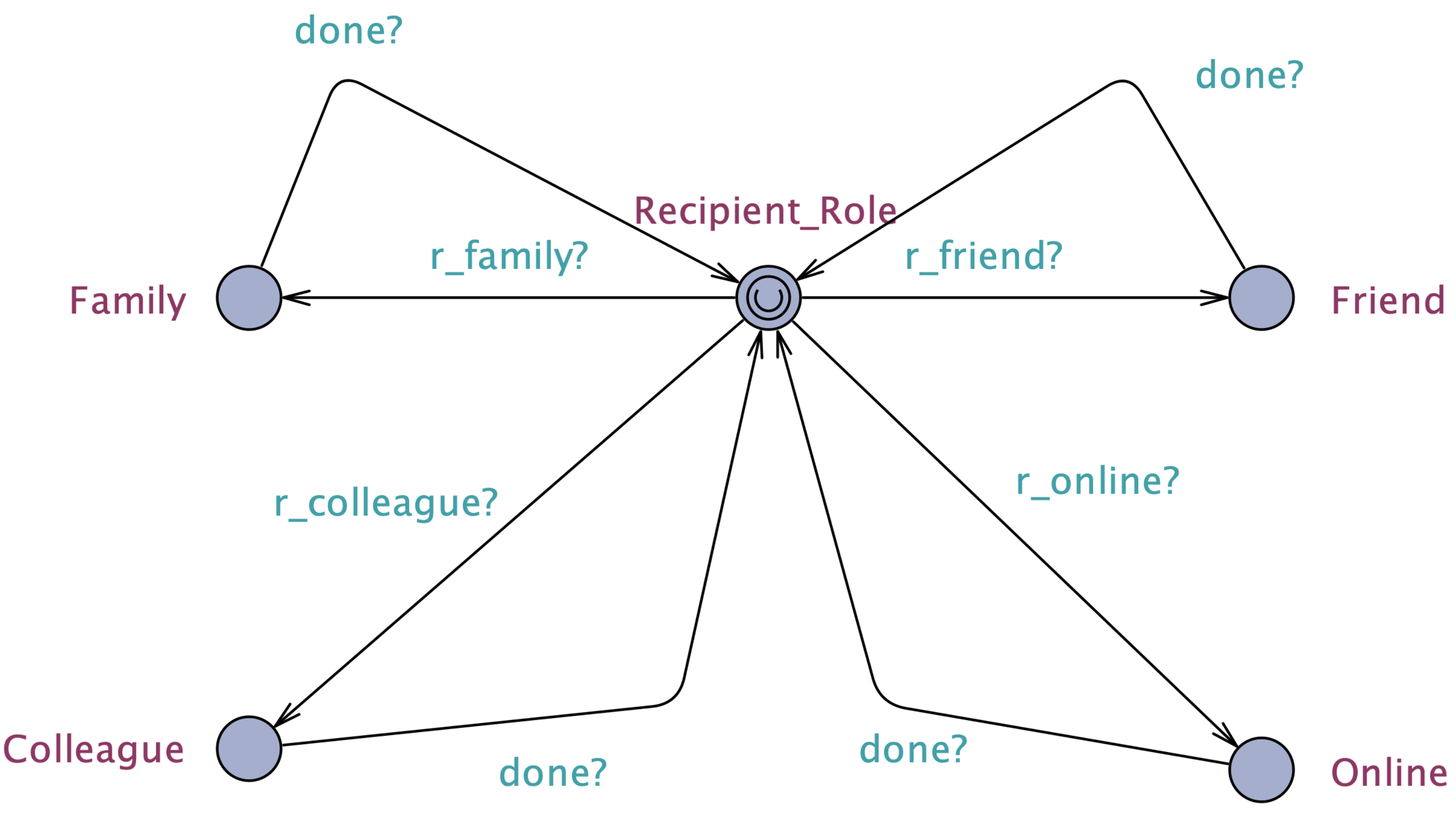}
         \caption{Recipient Role Observer}
         \label{fig:obs-rr}
     \end{subfigure}
        \caption{Observer Models Created in UPPAAL.}
        \label{fig:obs-models}
\end{figure*}

\subsection{Behavioral Analysis and Personalization}
To model the privacy disclosure behavior of a specific user, we collect the user's responses to the survey questionnaire and observe the information sharing behavior in different scenarios. For this, we randomly pick a user, for example, number 89 in our tabular dataset. Table \ref{tab:user-89-activities} contains the 8 random scenarios which were assigned to this user. Table \ref{tab:user-89-activities} represents that the user agreed to share information in 3 out of 8 given situations (scenario 1, 2, and 7). Based on that, we model the privacy behavior by composing them into a data dependent transition graph (Figure \ref{figure:user-1-fsm}). This graph contains a set of states and synchronization operations. When a transition happens from one state to another, a message is emitted to one or more observer processes through the synchronization channel. For example, when a transition happens from the \textit{Idle} state to state \textit{s1}, it emits a message titled \textit{finance} to any listening processes. This is one of the many useful features of UPPAAL which allows to design network of FSMs (i.e., parallel composition). The start and end states are marked as "committed states", which means there would be immediate transitions from these two states as soon as the transitions are enabled. In UPPAAL, the committed states take prompt transitions when the simulation or exhaustive search happens. This feature allows us to simulate the transitions spontaneously without waiting for any external inputs. It is worth mentioning that, we only model the positive sharing behavior of each user. In other words, figure \ref{figure:user-1-fsm} only contains a composition of 3 different scenarios where this user agreed to share the information with the recipients. Hence, if an information sharing attempt, described as a query, fails to comply with the model in figure \ref{figure:user-1-fsm}, then the model checker tells that the corresponding query was not satisfied and also shows a counter-example trace (if available).

\subsection{Observer Models}
An observer is an add-on automaton which without perturbing the observed system can detect events. We use 3 such models along with the user's behavioral model (Figure \ref{figure:user-1-fsm}) to keep track of the transitions and associated factors. This eventually help to prepare and employ descriptive queries for the verification of the model. Figure \ref{fig:obs-models} depicts those 3 separate observer models. Figure \ref{fig:obs-models} (a) represents the observer which keeps track of the information types. It listens for the messages--- \textit{finance}, \textit{health}, and \textit{relationship} whenever a transition in the behavioral model emits one of these values. For example, if a transition happens from \textit{Idle} state to the \textit{s1} state in the behavioral model (Figure \ref{figure:user-1-fsm}) then this observer model transitions from the \textit{Information\_Type} state to the \textit{Finance} state. The activities of the other two observer models (Figure \ref{fig:obs-models} (b) and (c)) are similar. Model \ref{fig:obs-models} (b) listens for the messages \textit{t\_family}, \textit{t\_friend}, \textit{t\_expert}, and \textit{t\_self} to keep track of the trust source. Likewise, model \ref{fig:obs-models} (c) listens for the messages \textit{r\_family}, \textit{r\_friend}, \textit{r\_colleague}, and \textit{r\_online} to keep track of the recipient's role. All the observer models return to their initial state once they get a specific message - \textit{done} from the behavioral model.

\subsection{Behavior as Systems}
The user-specific behavior model along with the observer models create the network automata otherwise known as a concurrent system in UPPAAL. This type of composition is also known as parallel composition of processes made of automaton. In our setup, the user model synchronizes data between itself and the observer models by leveraging the channel features in UPPAAL. The formal definition of the system model could be defines as follows:

$$User || Information\_Type || Trust\_Source || Recipient\_Role$$

\subsection{Validation}
\label{section:validation}
\begin{figure}
    \centering
    \includegraphics[width=1\linewidth]{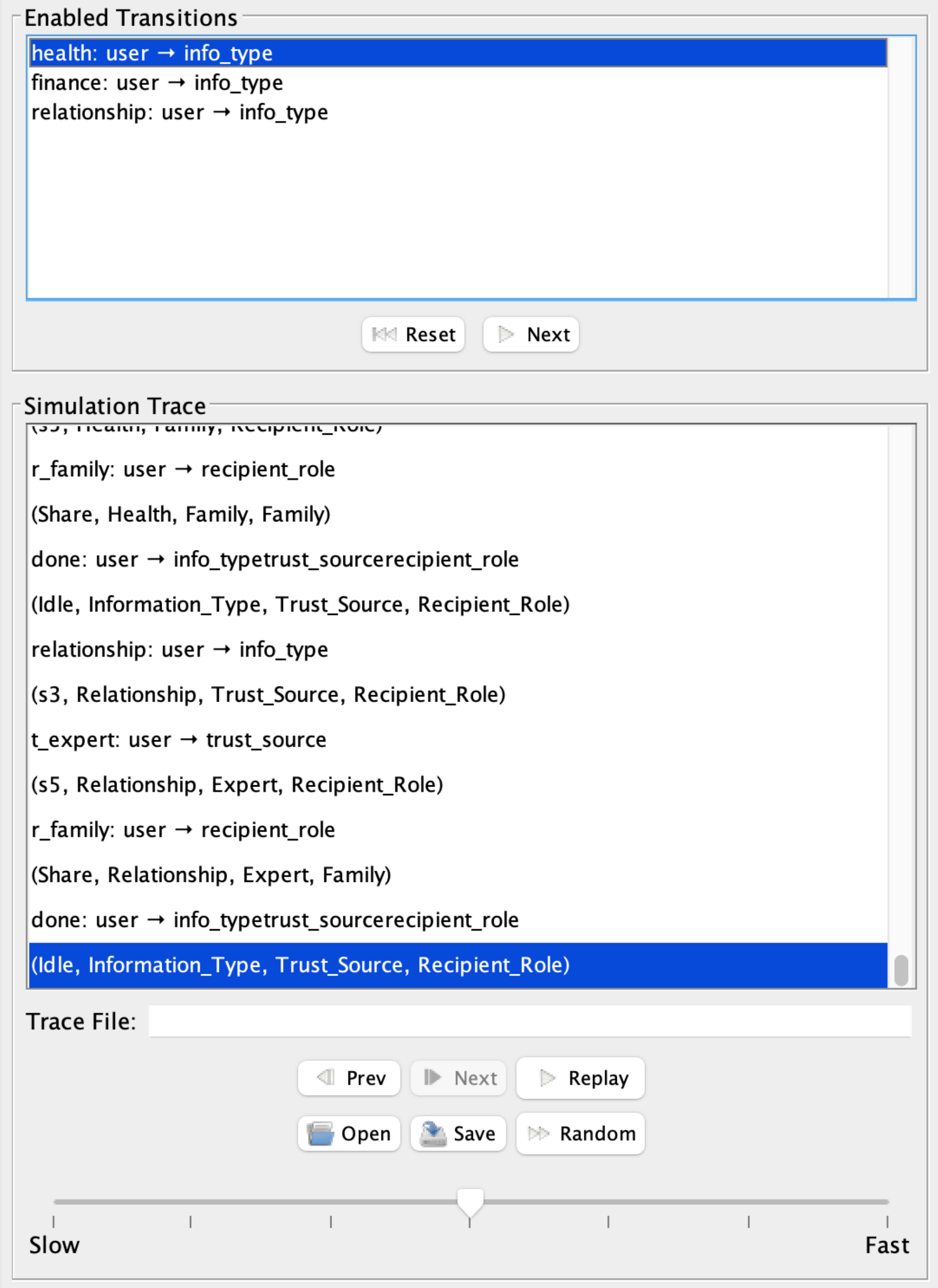}
    \caption{Part of the Simulation Window Containing the Control Buttons for Automatic and Manual Transition.}
    \Description{Description}
    \label{figure:simulation}
\end{figure}

UPPAAL uses graphical simulation as the model validation strategy \cite{behrmann2006tutorial}. Therefore, we conduct a simulation step to validate our models by running the system automatically which ensure that the models behave as intended, without any unexpected crash or deadlock. By utilizing the simulation feature of UPPAAL, we manually conduct some transitions in the behavioral model, and also utilize the random simulation feature to make sure the transitions are taken as expected. Figure \ref{figure:simulation} shows the UPPAAL simulation control panel where, the button \textit{Reset} and \textit{Next} are used to manually perform some transition operations, and the button \textit{Random} is used to start an automatic simulation that can run indefinitely. The simulation also allows us to make sure that the concurrency operation between the behavioral and the observer processes is taking place without any system breakdown.
 
\section{Verification with Model Checking}
\label{section:verification}
\begin{figure}
    \centering
    \includegraphics[width=0.75\linewidth]{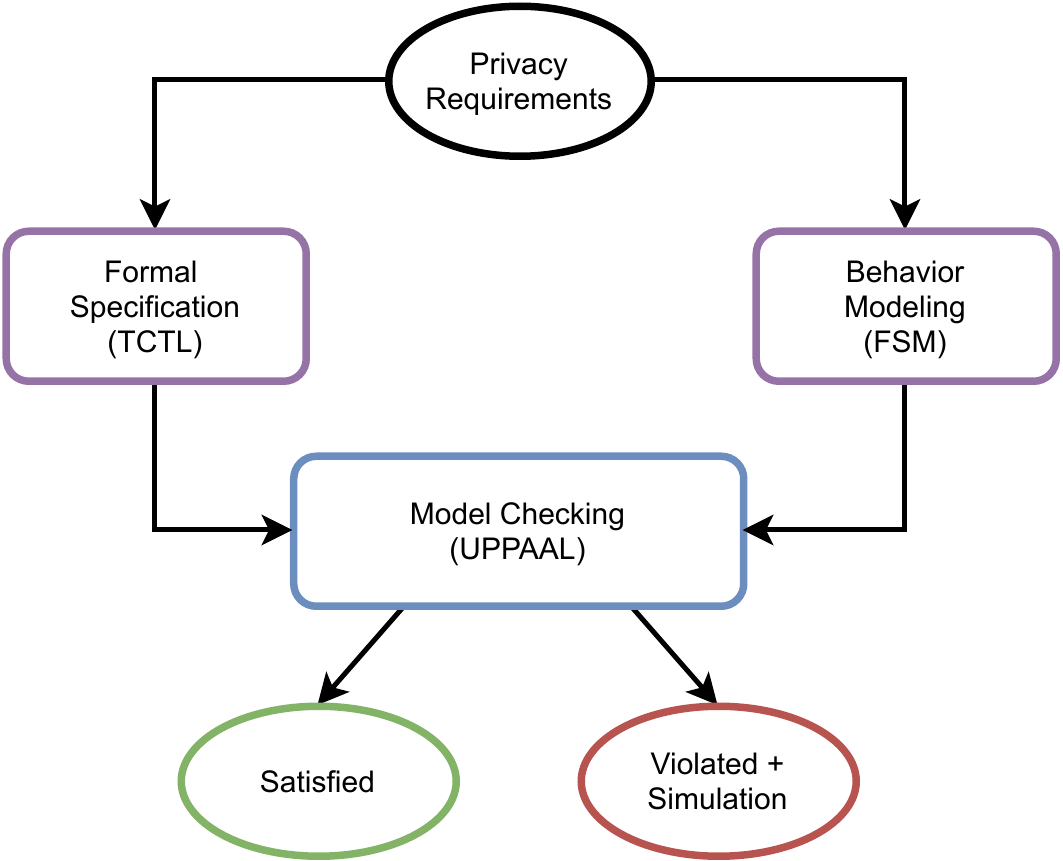}
    \caption{Model Checking Approach.}
    \Description{Description}
    \label{figure:model-checking}
\end{figure}

In this section, the verification of the user's privacy disclosure model is explained. Figure \ref{figure:model-checking} depicts a high level abstraction of the model checking process. In this approach, a set of desired properties (i.e., specifications) are checked against a model of a system\cite{bolton2014automatically,eleftherakis2001towards}.

\subsection{Specification Language in UPPAAL}
\label{section:privacy-properties}
The set of privacy properties (i.e., requirement specifications), which we expect the formal model to verify, are formulated based on the conducted survey in section 2.1. The specification languages that could be used to express these types of privacy properties are Linear Temporal Logic (LTL), Computation Tree Logic (CTL), and Timed Computation Tree Logic (TCTL)\cite{baier2008principles}. In UPPAAL, the process of verification operates with a simplified version of TCTL which is a subset of CTL. In TCTL, temporal connectiveness are expressed as pairs of symbols where the first element represents one of the path quantifiers and the second element represents one of the state quantifiers. Likewise, UPAAL query language consists of path formulae and state formulae \cite{behrmann2006tutorial}. The path formulae quantify over paths (traces) of the model whereas state formulae describe individual states. In UPPAAL, these quantifiers are expressed as follows:\\ 
$E$ = exists a path ($E$ in UPPAAL),\\
$A$ = for all paths ($A$ in UPPAAL),\\
$F$ = some state in a path ($<>$ in UPPAAL),\\
$G$ = all states in a path ($[]$ in UPPAAL),\\
Example queries could be written as $A[]p$, $A<>p$, $E<>p$, $E[]p$, and $p \rightarrow q$ where $p$ and $q$ are local properties. In other words, the query $E<>p$ tells that, 'it is possible to reach a state in which $p$ is satisfied' or '$p$ is true in at least one reachable state. $E<>Process.End$ is the UPPAAL notation for the same temporal logic formula $\exists\diamondsuit Process.End$ and is understood as 'it is possible to reach the location $End$ in automaton $Process$'.

\subsection{Personalized Privacy Properties}
In order to formulate the privacy properties of  user 89, we translate the user's disclosure decisions that are represented in Table ~\ref{tab:user-89-activities} in to the following statements:  'if the information type is \textit{health} and the trust source is a \textit{family} member and the recipient of the information is also a \textit{family} member, then the user \textit{share} the information. Similar to this specific criteria, every user has their own requirements when they agree to share the private information based on the situational factors. For each user, we translate their own privacy disclosure criteria into UPPAAL specification formulas. These formulas are then checked against his/her behavioral model to ensure the correctness of it. Since we use observer models (Figure \ref{fig:obs-models}) along with the behavioral model (Figure \ref{figure:user-1-fsm}) to create a concurrent system model, the observers have their own formal specification. In Table \ref{tab:properties}, we represent the equivalent expressions of the scenario factors in UPPAAL's specification language, while Figure \ref{fig:obs-models} visualizes the state transition graphs of those factors. Thus, the privacy disclosure properties for user 89 is represented in Table \ref{tab:user-properties} that is a the transformation of his/her responses based on the scenarios 1,2, and 3 from Table \ref{tab:user-89-activities}. Therefore, property number 1 from Table \ref{tab:user-properties} expresses: there exist a path, eventually where the properties enclosed in the parenthesis is true. 

\begin{table}
  \caption{Requirement Specifications or Privacy Properties of User 89}
  \label{tab:user-properties}
  \begin{tabular}{cl}
    \toprule
    No&Privacy Property\\
    \midrule
    1& \begin{tabular}[c]{@{}l@{}}E<> (user.share and information\_type.Health\\ and trust\_source.Family and recipient\_role.Family)\end{tabular}\\
    2& \begin{tabular}[c]{@{}l@{}}E<> (user.share and information\_type.Relationship\\ and trust\_source.Expert and recipient\_role.Family)\end{tabular}\\
    3& \begin{tabular}[c]{@{}l@{}}E<> (user.share and information\_type.Finance\\ and trust\_source.Expert and recipient\_role.Friend)\end{tabular}\\
  \bottomrule
\end{tabular}
\end{table}

\begin{table}
  \caption{Scenario Factors' Properties}
  \label{tab:properties}
  \begin{tabular}{cl}
    \toprule
    No&Knowledge Base Property\\
    \midrule
    1& $E<> (information\_type.Health)$\\
    2& $E<> (information\_type.Finance)$\\
    3& $E<> (information\_type.Relationship)$\\
    \midrule
    4& $E<> (trust\_source.Family)$\\
    5& $E<> (trust\_source.Friend)$\\
    6& $E<> (trust\_source.Expert)$\\
    7& $E<> (trust\_source.Self\_Search)$\\
    \midrule
    8& $E<> (recipient\_role.Family)$\\
    9& $E<> (recipient\_role.Friend)$\\
    10& $E<> (recipient\_role.Colleague)$\\
    12& $E<> (recipient\_role.Online\_Service)$\\
  \bottomrule
\end{tabular}
\end{table}

\subsection{Reachability Analysis}
There are three types of properties which are commonly checked against a formal model--- safety, liveness, and reachability properties. 
Reachability properties are used in state-transition systems which helps to examine the type and number of states that can be accessed through a particular system model \cite{kong2015dreach}. It is the simplest form of properties which determines whether a given state formula, $\Phi$, possibly could be satisfied by any reachable state. In this work, we verify whether or not the user-specific privacy properties holds in any, some, or all state of that user's privacy behavior model. We prefer reachability analysis over other similar methods (e.g., graph matching approach) because it allows us to search all potential paths in which the properties may or may not be satisfied, in a thorough and automated manner. Using UPPAAL, we applied reachability analysis to check which privacy properties were satisfied and which were not. UPPAAL performs the reachability analysis using either Breadth-First-Search or Depth-First-Search for checking whether a state is reachable or not. We preferred BFS of DFS to verify our reachability properties because it is a complete algorithm, ends within a finite time, and consider fewest edges while searching. The results of this procedure allows us to examine an user's privacy disclosure behavior, and whether or not a new sharing attempt complies with her existing privacy policies.

\begin{figure*}
    \centering
    \includegraphics[width=\linewidth]{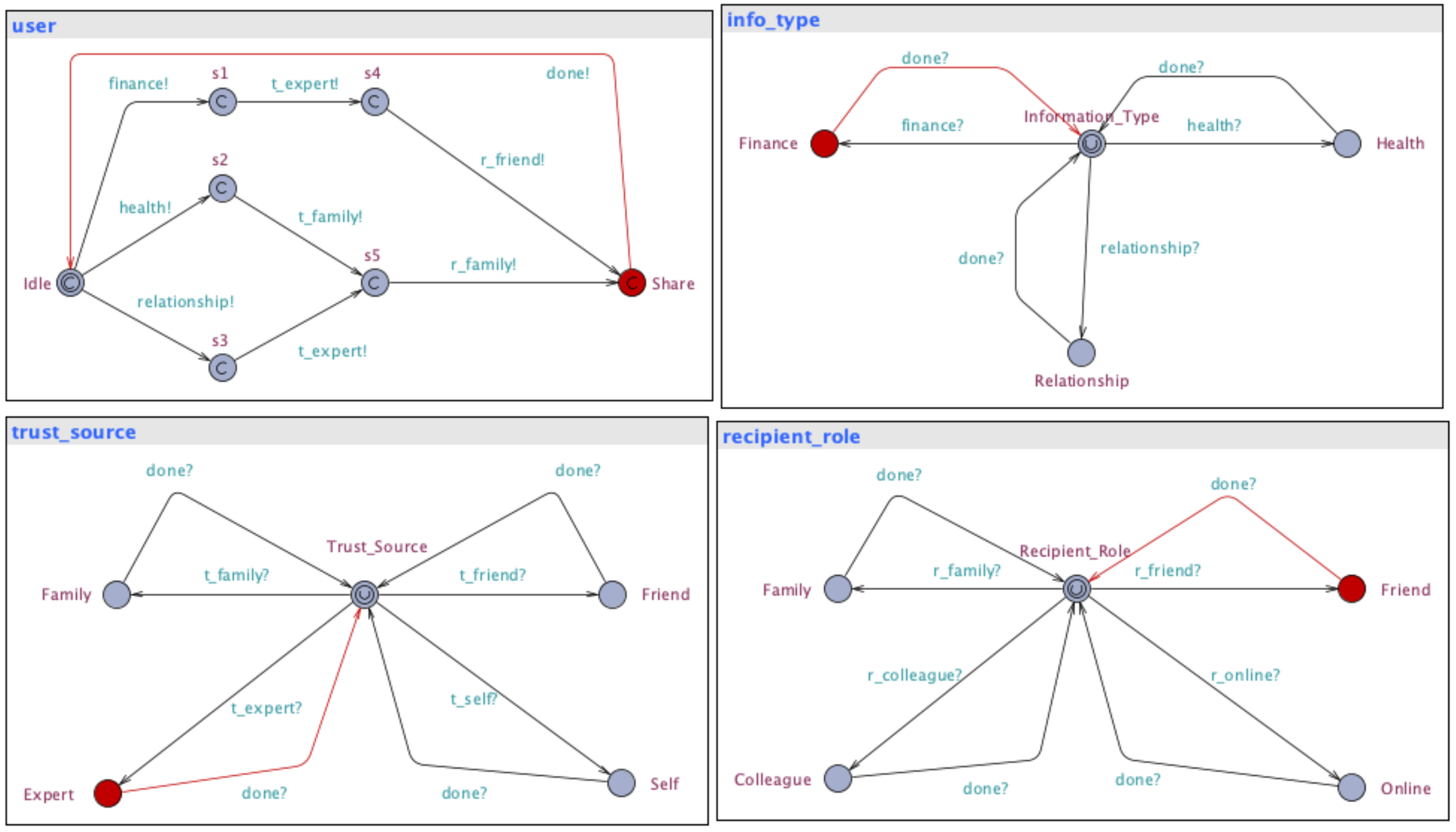}
    \caption{Diagnostic Trace of Query 3.}
    \Description{Description}
    \label{figure:trace}
\end{figure*}

Table \ref{tab:example-queries} contains a few verification queries that we check against the privacy disclosure model of user 89. Query 1 indeed gets satisfied since there is a valid transition in the FSM model (Idle -> s2 -> s5 -> Share) as well as in it's CTL version which is verified by the TCTL formula. Query 2 does not get satisfied since this user had no history of sharing his \textit{Health} information to either \textit{Friend} or \textit{Online} even when the trust source was \textit{Family}. Query 3 does not get satisfied because there is indeed one path where the property is true, (in Figure \ref{figure:user-1-fsm}, Idle -> s1 -> s4 -> Share). We can even see the diagnostic trace when this query is executed (Figure \ref{figure:trace}). Additionally, we can verify that the model will not face any deadlock in it's lifespan by executing queries like $\#$4.

\begin{figure}
    \centering
    \includegraphics[width=1\linewidth]{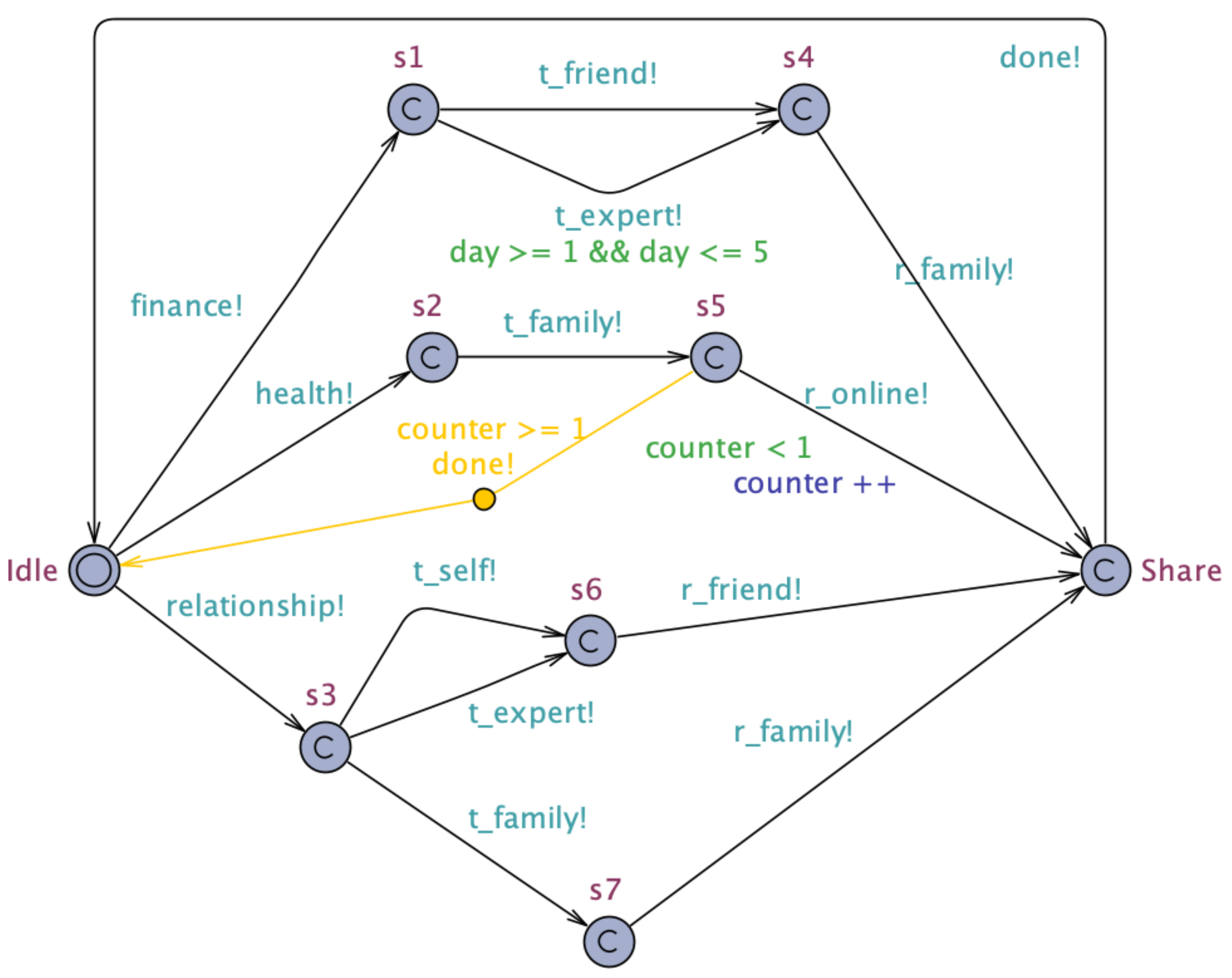}
    \caption{The Model of User 242 Created in UPPAAL.}
    \Description{Description}
    \label{figure:user-2-fsm}
    \vspace{-15pt}
\end{figure}

\begin{table*}
  \caption{Example of Some Queries to the User 89's Model and the Verification Results}
  \label{tab:example-queries}
  \begin{tabular}{lllll}
    \toprule
    No&To Verify&UPPAAL Query&Verification\\
    \midrule
    1 & \begin{tabular}[c]{@{}l@{}}There exists a path where eventually\\ the user is in \textit{Share} state and the information\\ type was \textit{Health}, trust source was \textit{Family},\\ and recipient's role was \textit{Family}\end{tabular} & \begin{tabular}[c]{@{}l@{}}E<>(user.Share and info\_type.Health and\\ trust\_source.Family and recipient\_role.Family)\end{tabular} & Satisfied\\
    \midrule
    2 & \begin{tabular}[c]{@{}l@{}}There exists a path where eventually\\ the user is in \textit{Share} state and the information\\ type was \textit{Health}, trust source was \textit{Family},\\ and recipient's role was either \textit{Friend} or \textit{Online}\end{tabular} & \begin{tabular}[c]{@{}l@{}}E<>(user.Share and info\_type.Health\\ and trust\_source.Family and\\ (recipient\_role.Friend\\ or recipient\_role.Online))\end{tabular} & Not Satisfied\\
    \midrule     
    3 & \begin{tabular}[c]{@{}l@{}}For all paths, it should never be the\\ case that the user is in \textit{Share} state and the\\ information type was \textit{Finance}, trust source\\ was \textit{Expert}, and recipient's role was \textit{Friend}\end{tabular} & \begin{tabular}[c]{@{}l@{}}A[] not (user.Share and info\_type.Finance\\ and trust\_source.Expert and\\ recipient\_role.Friend)\end{tabular} & Not Satisfied\\
    \midrule
    4 & \begin{tabular}[c]{@{}l@{}}There should not be any states\\ without successors\end{tabular} & \begin{tabular}[c]{@{}l@{}}E<> not deadlock\end{tabular} & Satisfied\\
  \bottomrule
\end{tabular}
\end{table*}

\section{Different Use Cases}
In this section, we represent the privacy disclosure model of a different user. A user is selected from the dataset randomly and holds the ID 242. In this case, in order to demonstrate the potential of the proposed behavioral model approach to include complex privacy properties with additional constraints, we imposed limitations on the days of the week or the number of times specific information could be disclosed. For this user, the responses that was received to the randomly assigned scenarios, we observed that this user agreed to share the information in 6 out of 8 situations. Therefore, we model his/her disclosure behavior in terms of a transition system (i.e., finite state machine) which is depicted in Figure \ref{figure:user-2-fsm}. As mentioned already, we added two guard conditions on two edges of the FSM: I) the day of the week for information sharing has to be between Monday to Friday (encoded as 1-5) to make the path- \textit{Idle -> Expert -> Family -> Share} enabled, II) \textit{Health} type information could be shared \textit{Online} no more than twice through the path \textit{Idle -> Health -> Family -> Online}. However, while verifying the model, we find a deadlock by querying \textit{E<> not deadlock} to the model checker. This property does not get satisfied depicting that there is indeed a deadlock. This happens because of the \textit{counter} guard which was imposed on the path, \textit{Idle -> Health -> Family -> Online}. Since a query is checked exhaustively, by running/simulating the model for hundreds of iterations, UPPAAL reaches to this deadlock state after simulating through this path for twice. In other words, the \textit{counter} become 1 and the path become disabled from the state \textit{s5} to \textit{Share}. However, we then resolve that deadlock in the model by creating a path from \textit{s5} to \textit{Idle} (colored in yellow). Thus, whenever the model checker tries to go through this path for more than twice and faces a guard in state \textit{s5}, it can then safely get back to the initial state without blocking the simulation operations.

In some other cases, incorporating additional decision-making factors or adding subcategories to the existing ones may result in a more complex network of automata with added granularity. For example, the information type \textit{health} could have two sub categories: mental health and physical health. A user might want to share \textit{physical health} condition with \textit{family} but \textit{mental health} condition to both \textit{family} and \textit{friends}. Representing this sort of scenario is also quite feasible in our proposed technique.

\subsection{Syntax and Semantics of the Models}
Each of the models in a system consists of a set of control nodes otherwise known as states. In addition to these control states, a composed model uses integer variables, simple channels, and broadcast channels. The edges of the automata contain two types of labels: guards and synchronization. The guards express the conditions on the values of the integer variables. These conditions need to be satisfied in order for the edges to be taken for transitions. In our models (e.g., Figure \ref{figure:user-2-fsm}), we add guards on transitions to ensure the traversal of the paths that represent the desired information sharing activity of the user. We also add synchronization variables in the models which enable the communication between the behavioral model and the observer models. In Figure \ref{figure:user-1-fsm} and Figure \ref{figure:user-2-fsm}, all the variables marked with an exclamation character "!" represent message transmission. Similarly, the observer models contain synchronization variables marked with a question mark "?" (Figure \ref{fig:obs-models}) that represent message reception. For example, whenever a transition happens from the state \textit{Idle} to the state \textit{s1}, on the behavioral model (Figure \ref{figure:user-1-fsm}), it transmits a message \textit{finance} which is then received by the "Information Type" observer model through the \textit{finance?} path. The simple channels (e.g, finance!, finance? t\_family!, r\_friend!) help the observer models to keep track of the scenario factors and the broadcasting channel (i.e., \textit{done!}) helps the observers to get back to the start position once an iteration (i.e., sharing activity) is completed. The model also consists of urgent (as soon as transition is enabled, current state will change to the next state) and committed locations. Since the information sharing behavior is assumed to be an non time-dependent process, we make the states urgent so that the transitions happen as soon as the flags are available. Its worth mentioning that, in time-dependent systems, the use of urgent locations reduces the complexity of the analysis by reducing the number of clocks.

\section{Application and Usability}
\label{section:usability}
In this section, we briefly describe the application and usability of our proposed methodology. In the following section, we describe a process of creating the baseline model from user's historical sharing activities. 

\subsection{Automatic Translation of Activities to DFA}
Formal modeling, formal verification and validation approaches are mostly used in the area of physical systems (e.g., industrial control systems, and cyber-physical systems). In this context, the process is mostly completed by human domain experts. Once a model is developed, validated, and verified; it is then used as the foundation of any downstream tasks such as hardware assembly, resiliency test, etc. In contrast, in the user-specific behavioral study, the formal modeling part has to be a automatic process that translated user's desired privacy properties into formal specifications. This is because the end users of an application will not have sufficient expertise to create the mathematically based model which is a core requirement to the model checking technique. Therefore, we utilize an existing tool \cite{noam} to automatically translate an user's historical sharing activities, manifested as regular expression, into deterministic finite automata (DFA).

\begin{figure}
    \centering
    \includegraphics[width=0.75\linewidth]{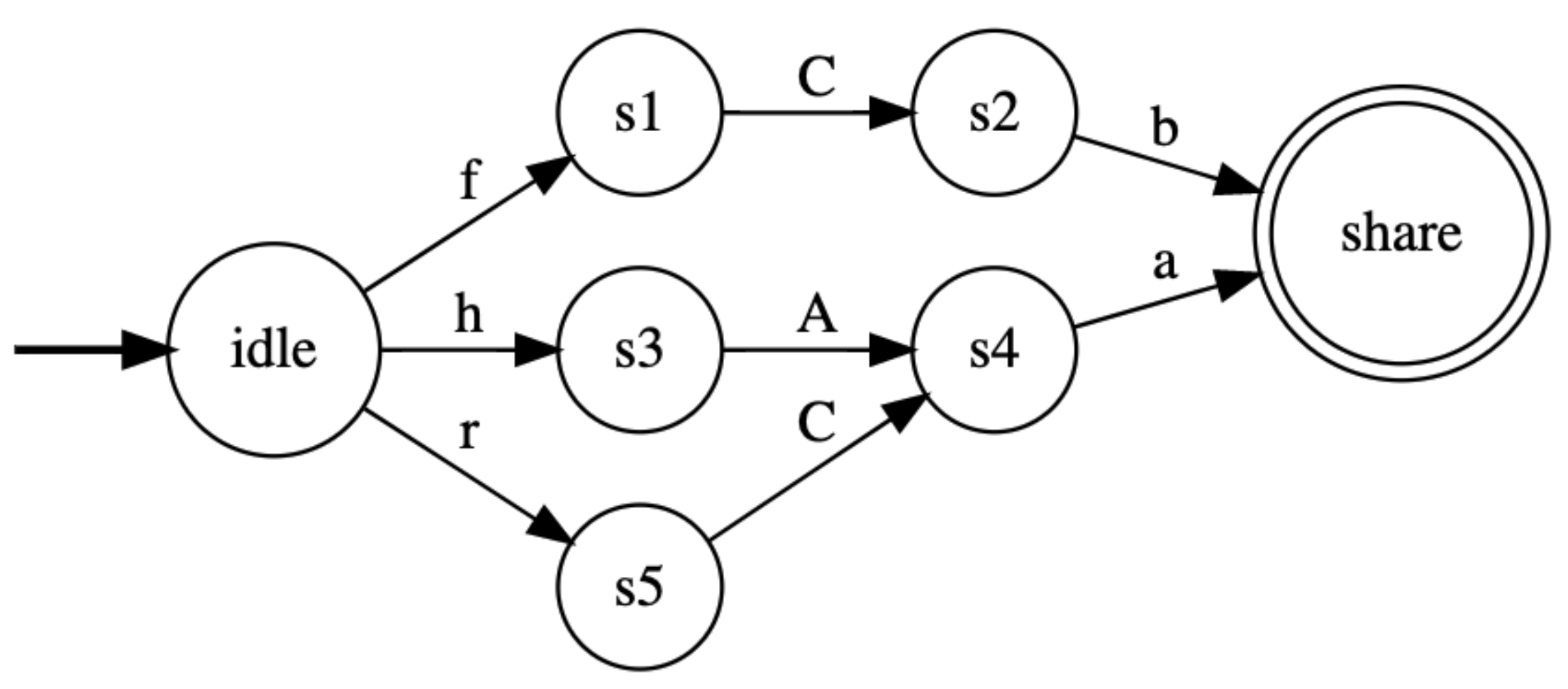}
    \caption{The DFA of the User 89.}
    \Description{Description}
    \label{figure:user-1-dfa}
      \vspace{-15pt}
\end{figure}

First, we generate a regular expression string for every shared activity. We get three such strings--- \textit{rCa}, \textit{hAa}, and \textit{fCb} from the sharing activity (i.e., survey response) of the user 89 (Table \ref{tab:user-89-activities}). Where, $h$,$f$, and $r$ represent the information types- $Health$, $Finance$, $Relationship$ respectively. $A$,$B$, $C$, and $D$ represent the trust source--- $Family$, $Friend$, $Expert$, and $Self Search$ respectively. $a$,$b$, $c$, and $d$ represent the recipient's role--- $Family$, $Friend$, $Colleague$, and $Online Service$ respectively. Then we combine the strings together with the regular expression's choice character 
$+$ to get $$(rC+hA)a+fCb$$ 
Finally, we use the tool \cite{noam} to generate a minimized DFA that accepts the regular expression (i.e., model the shared activities in terms of finite automata). Below is the formal Definition of the DFA:
$$\text{Set of state, }Q = \{idle,s_1,s_2,s_3,s_4,s_5,Share\}$$
$$\text{Alphabet, }\Sigma = \{h,f,r,A,B,C,D,a,b,c,d\}$$
$$\text{Initial state, }q_o = idle$$
$$\text{Set of final states, }F = \{share\}$$
$$\text{Transition function, }\delta = Q \times \Sigma \rightarrow Q$$
\\
Figure \ref{figure:user-1-dfa} is the result of the translation which acts as the foundation of the UPPAAL model depicted in Figure \ref{figure:user-1-fsm}. This preliminary automation step could be later taken over by another downstream automation tool (discussed later) to eventually develop a UPPAAL acceptable formal model.
%

\subsection{Standalone Privacy Management Tool}
Normally, the verification engine of UPPAAL is by default executed on the same computer as the user interface, but it can also run on a more powerful server which allows to host a complex behavioral model. Another supporting utility named \textit{verifyta} is able to accept \textit{.ta}, \textit{.xta}, and \textit{.xml} files as an input and use high-level programming language (e.g., Java)) API to perform the modeling, simulation, and model checking through a pragmatically native environment. This API makes it possible to interpret user's historical sharing activities and then develop a UPPAAL compatible formal model. The API could additionally be utilized to validate the developing model and verify it against a set of queries.

\subsection{In Software Design and Development}
One of the many advantages of formal modeling is its ability to allow for an early assessment of the model \cite{yoo2009formal, alagar2011specification}. In other words, it is possible to design, validate, and exhaustively verify user's privacy behavior model and think of it as the algorithm of his allowed behavior. Later on, programmers can leverage this model as the template for coding a function (e.g., shouldShare()) for that user in their software system. This process will enable the programmers to write a function which is already exhaustively tested, and therefore, no need to conduct typical unit testing on the program. An existing software can also integrate the privacy management tool by interacting with the verification engine through high level API. Thus the software can achieve a proper privacy management component inside it's ecosystem.

\subsection{User-Interface (UI) of the Privacy Settings}
The user-interface of any software, mobile-app, or web-app plays an important role in providing its users with a more flexible privacy settings. Users of the communication platforms are found to be less careful about properly setting their privacy preferences offered by the apps \cite{mazzia2012pviz, lipford2008understanding, liu2011analyzing}. This is because of the generic and 'one fits all' nature of the privacy preference pages. Therefore, user-specific formal modeling can help with the UI/UX designers and programmers are better equipped to provide personalized privacy settings pages to their users by utilizing their underlying behavioral model. 
  \vspace{-10pt}
\section{Related Work}
\label{section:related-work}
Researchers from the field of privacy, decision making, and personalization have shed light on the area of behavior modeling. They have been exploring, how the psychological factors of humans relate to their concerns about their information privacy \cite{westin1968privacy, ackerman1999privacy, lipford2008understanding}. Accordingly, many behavioral theories have been established and adopted to the privacy management domain \cite{ajzen1991theory, barth2006privacy, hale2002theory}. Theory of planned behavior (TPB) tells that people's behavior is directly determined by their behavioral \textit{intentions}. These intentions are in turn influenced by their \textit{attitude} (positive or negative evaluation of the decision), perception of the \textit{subjective norms} (generally expected behavior in their social group), and \textit{perceived behavioral control} (ease or difficulty to perform the behavior). The theory also states that these constructs together determine an individual's behavioral intentions and provide a model to capture humans' decision making behavior. Therefore, researchers from various areas (e.g., privacy, use of the internet, health, environmental psychology, etc) have used TPB and demonstrated its effectiveness in predicting human behavior in terms of privacy decision making\cite{heirman2013predicting, van1999involving, yao2008predicting, conner2003environmental, macovei2015determinants}. 

Another privacy management theory which is relevant to our work is known as the theory of contextual integrity (CI) \cite{barth2006privacy}. In the CI theory, privacy is formulated as an appropriate flow of information that conforms with the contextual informational norms (i.e., rules governing flow of information in CI format). An example of a norm in the context of \textit{health} could be: a husband usually shares his diagnosis result with his family doctor, or his wife but not with his friends or in the social media. In this example, the husband is recognized as the data subject and the sender, the doctor or wife as the recipient of the information, health as the information type, and the recipient will hold the information confidentially as the transmission principle. Based on the theory of (CI) \cite{barth2006privacy}, privacy is violated if the information is shared or transferred with friends or financial advisers, as they are not usually and explicitly included as part of the 'allowed' recipients of the information. 

Consequently, many researchers have studied modeling users' privacy decision-making process in the context of various types and recipients of the information. Knijnenburg et. al. discovered about how the type of the information and their recipients have significant effect on user's information disclosing tendency \cite{knijnenburg2014increasing}. In their study participants were asked to set their privacy settings on a custom made privacy settings UI of an imagined Facebook-like social network site by indicating which of their profile information they would share with whom. In another study \cite{dinev2006extended}, authors have examined the idea of users' privacy calculus (i.e., costs vs benefits) and how it led the users to disclose their different types of private information to different types of recipients (websites), in a purpose-specific fashion. Lederer et. al. \cite{lederer2003wants} investigated the relative effects of different recipients and the situations towards users' information disclosure intention. By surveying 130 participants, given two hypothetical situations, they found that situation is an important determinant and highly correlated with the information recipient.

Despite the existence of many behavioral theories and analysis, only a handful of works address the issue of personalized modeling of human behavior. Most importantly, a few of them acknowledge the issue of practical usability and application of the derived models. Joshaghani et. al. extends the concept of CI theory and provide mathematical models that enables the creations and management of privacy norms by the individual users \cite{joshaghani2019formal}. They propose and develop a custom formal verification technique which ensure privacy norms are enforced for every information sharing attempt by the user. Similar to our transition system based formalism, Lu et. al. proposed a technique that translates the privacy specification or requirements of web services to LTL formulas \cite{lu2014verification}. Then the create the privacy policy model by utilizing a privacy interface automata (PIA) that transforms the messaging structure extracted from the web service business process execution language into an automaton. Krishnan et al. propose a semi-formal approach that enforce privacy requirements by leveraging the role-based access control technique along with LTL formulas \cite{krishnan2013enforcement}. Grace et al. propose a technique for modeling user-centric privacy management using labeled transition systems. The goal of this model is to compare the user's privacy preferences with the privacy policies of the cloud service provider \cite{grace2017towards}. Thus the users 'can be informed of the privacy implications of the services' and warned of potential privacy breaches. However, they mentioned about two limitations--- i) requirement of human intervention for creating initial model, ii) limited extensibility and scalability.

In our work, we address many of the above-mentioned limitations and open questions by representing personalized situational behavior, proposing a technique for automatic translation of activities to FSM, demonstrating the practical usability, and describing the scalability of this formal approach. 

\section{Conclusion}
\label{section:conclusion}
Users' ability to better manage their data-sharing practices is limited due to the lack of suitable user-centric privacy management tools and techniques. Moreover, very few of the existing methodologies take into consideration the aspect of personalization, correctness, and explainability. Most importantly, their practical usability and acceptance remain a significant challenge. In this paper we have presented an approach to formally model, validate, and verify personalized privacy disclosure behavior based on the analysis of user's situational decision-making process. The proposed methodology demonstrates a privacy formalism and verification technique based on UPPAAL which is a tool for modeling, validation, and verification of automata based systems. Most importantly, the methodology depicts the potential of formalism towards the development of user-centric privacy management tools. In future work, we plan to extend the user's privacy behavior model to incorporate additional decision making factors towards more granularity. We also plan to develop an end-to-end framework on top of UPPAAL to fully automate the process of transforming the historical sharing activities into UPPAAL compatible network of automata.

\begin{acks}
The authors would like to thank National Science Foundation for
its support through the Computer and Information Science and
Engineering (CISE) program and Research Initiation Initiative(CRII)
grant number 1657774 of the Secure and Trustworthy Cyberspace
(SaTC) program: A System for Privacy Management in Ubiquitous
Environments.
\end{acks}
\vspace{-10pt}
\bibliographystyle{ACM-Reference-Format}
\bibliography{main}


\begin{thebibliography}{54}


\ifx \showCODEN    \undefined \def \showCODEN     #1{\unskip}     \fi
\ifx \showDOI      \undefined \def \showDOI       #1{#1}\fi
\ifx \showISBNx    \undefined \def \showISBNx     #1{\unskip}     \fi
\ifx \showISBNxiii \undefined \def \showISBNxiii  #1{\unskip}     \fi
\ifx \showISSN     \undefined \def \showISSN      #1{\unskip}     \fi
\ifx \showLCCN     \undefined \def \showLCCN      #1{\unskip}     \fi
\ifx \shownote     \undefined \def \shownote      #1{#1}          \fi
\ifx \showarticletitle \undefined \def \showarticletitle #1{#1}   \fi
\ifx \showURL      \undefined \def \showURL       {\relax}        \fi
\providecommand\bibfield[2]{#2}
\providecommand\bibinfo[2]{#2}
\providecommand\natexlab[1]{#1}
\providecommand\showeprint[2][]{arXiv:#2}

\bibitem[\protect\citeauthoryear{Ackerman, Cranor, and Reagle}{Ackerman
  et~al\mbox{.}}{1999}]%
        {ackerman1999privacy}
\bibfield{author}{\bibinfo{person}{Mark~S Ackerman},
  \bibinfo{person}{Lorrie~Faith Cranor}, {and} \bibinfo{person}{Joseph
  Reagle}.} \bibinfo{year}{1999}\natexlab{}.
\newblock \showarticletitle{Privacy in e-commerce: examining user scenarios and
  privacy preferences}. In \bibinfo{booktitle}{\emph{Proceedings of the 1st ACM
  conference on Electronic commerce}}. \bibinfo{pages}{1--8}.
\newblock


\bibitem[\protect\citeauthoryear{Afifi and Guerrero}{Afifi and
  Guerrero}{2000}]%
        {afifi2000motivations}
\bibfield{author}{\bibinfo{person}{Walid~A Afifi} {and}
  \bibinfo{person}{Laura~K Guerrero}.} \bibinfo{year}{2000}\natexlab{}.
\newblock \showarticletitle{Motivations underlying topic avoidance in close
  relationships}.
\newblock \bibinfo{journal}{\emph{Balancing the secrets of private
  disclosures}} (\bibinfo{year}{2000}), \bibinfo{pages}{165--180}.
\newblock


\bibitem[\protect\citeauthoryear{Ajzen et~al\mbox{.}}{Ajzen
  et~al\mbox{.}}{1991}]%
        {ajzen1991theory}
\bibfield{author}{\bibinfo{person}{Icek Ajzen} {et~al\mbox{.}}}
  \bibinfo{year}{1991}\natexlab{}.
\newblock \showarticletitle{The theory of planned behavior}.
\newblock \bibinfo{journal}{\emph{Organizational behavior and human decision
  processes}} \bibinfo{volume}{50}, \bibinfo{number}{2} (\bibinfo{year}{1991}),
  \bibinfo{pages}{179--211}.
\newblock


\bibitem[\protect\citeauthoryear{Alagar and Periyasamy}{Alagar and
  Periyasamy}{2011}]%
        {alagar2011specification}
\bibfield{author}{\bibinfo{person}{Vangalur~S Alagar} {and}
  \bibinfo{person}{Kasilingam Periyasamy}.} \bibinfo{year}{2011}\natexlab{}.
\newblock \bibinfo{booktitle}{\emph{Specification of software systems}}.
\newblock \bibinfo{publisher}{Springer Science \& Business Media}.
\newblock


\bibitem[\protect\citeauthoryear{Ashley, Hada, Karjoth, Powers, and
  Schunter}{Ashley et~al\mbox{.}}{2003}]%
        {ashley2003enterprise}
\bibfield{author}{\bibinfo{person}{Paul Ashley}, \bibinfo{person}{Satoshi
  Hada}, \bibinfo{person}{G{\"u}nter Karjoth}, \bibinfo{person}{Calvin Powers},
  {and} \bibinfo{person}{Matthias Schunter}.} \bibinfo{year}{2003}\natexlab{}.
\newblock \showarticletitle{Enterprise privacy authorization language (EPAL)}.
\newblock \bibinfo{journal}{\emph{IBM Research}}  \bibinfo{volume}{30}
  (\bibinfo{year}{2003}), \bibinfo{pages}{31}.
\newblock


\bibitem[\protect\citeauthoryear{Aucher, Boella, and Van Der~Torre}{Aucher
  et~al\mbox{.}}{2011}]%
        {aucher2011dynamic}
\bibfield{author}{\bibinfo{person}{Guillaume Aucher}, \bibinfo{person}{Guido
  Boella}, {and} \bibinfo{person}{Leendert Van Der~Torre}.}
  \bibinfo{year}{2011}\natexlab{}.
\newblock \showarticletitle{A dynamic logic for privacy compliance}.
\newblock \bibinfo{journal}{\emph{Artificial Intelligence and Law}}
  \bibinfo{volume}{19}, \bibinfo{number}{2-3} (\bibinfo{year}{2011}),
  \bibinfo{pages}{187}.
\newblock


\bibitem[\protect\citeauthoryear{Baier and Katoen}{Baier and Katoen}{2008}]%
        {baier2008principles}
\bibfield{author}{\bibinfo{person}{Christel Baier} {and}
  \bibinfo{person}{Joost-Pieter Katoen}.} \bibinfo{year}{2008}\natexlab{}.
\newblock \bibinfo{booktitle}{\emph{Principles of model checking}}.
\newblock \bibinfo{publisher}{MIT press}.
\newblock


\bibitem[\protect\citeauthoryear{Barth, Datta, Mitchell, and Nissenbaum}{Barth
  et~al\mbox{.}}{2006}]%
        {barth2006privacy}
\bibfield{author}{\bibinfo{person}{Adam Barth}, \bibinfo{person}{Anupam Datta},
  \bibinfo{person}{John~C Mitchell}, {and} \bibinfo{person}{Helen Nissenbaum}.}
  \bibinfo{year}{2006}\natexlab{}.
\newblock \showarticletitle{Privacy and contextual integrity: Framework and
  applications}. In \bibinfo{booktitle}{\emph{2006 IEEE Symposium on Security
  and Privacy (S\&P'06)}}. IEEE, \bibinfo{pages}{15--pp}.
\newblock


\bibitem[\protect\citeauthoryear{Behrmann, David, and Larsen}{Behrmann
  et~al\mbox{.}}{2006}]%
        {behrmann2006tutorial}
\bibfield{author}{\bibinfo{person}{Gerd Behrmann}, \bibinfo{person}{Alexandre
  David}, {and} \bibinfo{person}{Kim~G Larsen}.}
  \bibinfo{year}{2006}\natexlab{}.
\newblock \showarticletitle{A tutorial on Uppaal 4.0}.
\newblock \bibinfo{journal}{\emph{Department of computer science, Aalborg
  university}} (\bibinfo{year}{2006}).
\newblock


\bibitem[\protect\citeauthoryear{Bolton, Jim{\'e}nez, van Paassen, and
  Trujillo}{Bolton et~al\mbox{.}}{2014}]%
        {bolton2014automatically}
\bibfield{author}{\bibinfo{person}{Matthew~L Bolton}, \bibinfo{person}{Noelia
  Jim{\'e}nez}, \bibinfo{person}{Marinus~M van Paassen}, {and}
  \bibinfo{person}{Maite Trujillo}.} \bibinfo{year}{2014}\natexlab{}.
\newblock \showarticletitle{Automatically generating specification properties
  from task models for the formal verification of human--automation
  interaction}.
\newblock \bibinfo{journal}{\emph{IEEE Transactions on Human-Machine Systems}}
  \bibinfo{volume}{44}, \bibinfo{number}{5} (\bibinfo{year}{2014}),
  \bibinfo{pages}{561--575}.
\newblock


\bibitem[\protect\citeauthoryear{Breaux, Hibshi, and Rao}{Breaux
  et~al\mbox{.}}{2014}]%
        {breaux2014eddy}
\bibfield{author}{\bibinfo{person}{Travis~D Breaux}, \bibinfo{person}{Hanan
  Hibshi}, {and} \bibinfo{person}{Ashwini Rao}.}
  \bibinfo{year}{2014}\natexlab{}.
\newblock \showarticletitle{Eddy, a formal language for specifying and
  analyzing data flow specifications for conflicting privacy requirements}.
\newblock \bibinfo{journal}{\emph{Requirements Engineering}}
  \bibinfo{volume}{19}, \bibinfo{number}{3} (\bibinfo{year}{2014}),
  \bibinfo{pages}{281--307}.
\newblock


\bibitem[\protect\citeauthoryear{Buchanan, Paine, Joinson, and Reips}{Buchanan
  et~al\mbox{.}}{2007}]%
        {buchanan2007development}
\bibfield{author}{\bibinfo{person}{Tom Buchanan}, \bibinfo{person}{Carina
  Paine}, \bibinfo{person}{Adam~N Joinson}, {and} \bibinfo{person}{Ulf-Dietrich
  Reips}.} \bibinfo{year}{2007}\natexlab{}.
\newblock \showarticletitle{Development of measures of online privacy concern
  and protection for use on the Internet}.
\newblock \bibinfo{journal}{\emph{Journal of the Association for Information
  Science and Technology}} \bibinfo{volume}{58}, \bibinfo{number}{2}
  (\bibinfo{year}{2007}), \bibinfo{pages}{157--165}.
\newblock


\bibitem[\protect\citeauthoryear{Clarke and Wing}{Clarke and Wing}{1996}]%
        {clarke1996formal}
\bibfield{author}{\bibinfo{person}{Edmund~M Clarke} {and}
  \bibinfo{person}{Jeannette~M Wing}.} \bibinfo{year}{1996}\natexlab{}.
\newblock \showarticletitle{Formal methods: State of the art and future
  directions}.
\newblock \bibinfo{journal}{\emph{ACM Computing Surveys (CSUR)}}
  \bibinfo{volume}{28}, \bibinfo{number}{4} (\bibinfo{year}{1996}),
  \bibinfo{pages}{626--643}.
\newblock


\bibitem[\protect\citeauthoryear{Conner, Kirk, Cade, and Barrett}{Conner
  et~al\mbox{.}}{2003}]%
        {conner2003environmental}
\bibfield{author}{\bibinfo{person}{Mark Conner}, \bibinfo{person}{Sara~FL
  Kirk}, \bibinfo{person}{Janet~E Cade}, {and} \bibinfo{person}{Jennifer~H
  Barrett}.} \bibinfo{year}{2003}\natexlab{}.
\newblock \showarticletitle{Environmental influences: factors influencing a
  woman's decision to use dietary supplements}.
\newblock \bibinfo{journal}{\emph{The Journal of nutrition}}
  \bibinfo{volume}{133}, \bibinfo{number}{6} (\bibinfo{year}{2003}),
  \bibinfo{pages}{1978S--1982S}.
\newblock


\bibitem[\protect\citeauthoryear{Costante, Sun, Petkovi{\'c}, and
  Den~Hartog}{Costante et~al\mbox{.}}{2012}]%
        {costante2012machine}
\bibfield{author}{\bibinfo{person}{Elisa Costante}, \bibinfo{person}{Yuanhao
  Sun}, \bibinfo{person}{Milan Petkovi{\'c}}, {and} \bibinfo{person}{Jerry
  Den~Hartog}.} \bibinfo{year}{2012}\natexlab{}.
\newblock \showarticletitle{A machine learning solution to assess privacy
  policy completeness: (short paper)}. In \bibinfo{booktitle}{\emph{Proceedings
  of the 2012 ACM Workshop on Privacy in the Electronic Society}}.
  \bibinfo{pages}{91--96}.
\newblock


\bibitem[\protect\citeauthoryear{Cranor}{Cranor}{2002}]%
        {cranor2002web}
\bibfield{author}{\bibinfo{person}{Lorrie Cranor}.}
  \bibinfo{year}{2002}\natexlab{}.
\newblock \bibinfo{booktitle}{\emph{Web privacy with P3P}}.
\newblock \bibinfo{publisher}{" O'Reilly Media, Inc."}.
\newblock


\bibitem[\protect\citeauthoryear{Dinev and Hart}{Dinev and Hart}{2006}]%
        {dinev2006extended}
\bibfield{author}{\bibinfo{person}{Tamara Dinev} {and} \bibinfo{person}{Paul
  Hart}.} \bibinfo{year}{2006}\natexlab{}.
\newblock \showarticletitle{An extended privacy calculus model for e-commerce
  transactions}.
\newblock \bibinfo{journal}{\emph{Information systems research}}
  \bibinfo{volume}{17}, \bibinfo{number}{1} (\bibinfo{year}{2006}),
  \bibinfo{pages}{61--80}.
\newblock


\bibitem[\protect\citeauthoryear{Eleftherakis and Kefalas}{Eleftherakis and
  Kefalas}{2001}]%
        {eleftherakis2001towards}
\bibfield{author}{\bibinfo{person}{G Eleftherakis} {and} \bibinfo{person}{P
  Kefalas}.} \bibinfo{year}{2001}\natexlab{}.
\newblock \showarticletitle{Towards model checking of finite state machines
  extended with memory through refinement}.
\newblock \bibinfo{journal}{\emph{Advances in signal processing and computer
  technologies}} (\bibinfo{year}{2001}), \bibinfo{pages}{321--326}.
\newblock


\bibitem[\protect\citeauthoryear{Grace and Surridge}{Grace and
  Surridge}{2017}]%
        {grace2017towards}
\bibfield{author}{\bibinfo{person}{Paul Grace} {and} \bibinfo{person}{Mike
  Surridge}.} \bibinfo{year}{2017}\natexlab{}.
\newblock \showarticletitle{Towards a model of user-centered privacy
  preservation}. In \bibinfo{booktitle}{\emph{Proceedings of the 12th
  International Conference on Availability, Reliability and Security}}.
  \bibinfo{pages}{1--8}.
\newblock


\bibitem[\protect\citeauthoryear{Grumberg, Peled, and Clarke}{Grumberg
  et~al\mbox{.}}{1999}]%
        {grumberg1999model}
\bibfield{author}{\bibinfo{person}{Orna Grumberg}, \bibinfo{person}{Doron~A
  Peled}, {and} \bibinfo{person}{EM Clarke}.} \bibinfo{year}{1999}\natexlab{}.
\newblock \bibinfo{title}{Model checking}.
\newblock
\newblock


\bibitem[\protect\citeauthoryear{Hale, Householder, and Greene}{Hale
  et~al\mbox{.}}{2002}]%
        {hale2002theory}
\bibfield{author}{\bibinfo{person}{Jerold~L Hale}, \bibinfo{person}{Brian~J
  Householder}, {and} \bibinfo{person}{Kathryn~L Greene}.}
  \bibinfo{year}{2002}\natexlab{}.
\newblock \showarticletitle{The theory of reasoned action}.
\newblock \bibinfo{journal}{\emph{The persuasion handbook: Developments in
  theory and practice}}  \bibinfo{volume}{14} (\bibinfo{year}{2002}),
  \bibinfo{pages}{259--286}.
\newblock


\bibitem[\protect\citeauthoryear{Heirman, Walrave, and Ponnet}{Heirman
  et~al\mbox{.}}{2013}]%
        {heirman2013predicting}
\bibfield{author}{\bibinfo{person}{Wannes Heirman}, \bibinfo{person}{Michel
  Walrave}, {and} \bibinfo{person}{Koen Ponnet}.}
  \bibinfo{year}{2013}\natexlab{}.
\newblock \showarticletitle{Predicting adolescents' disclosure of personal
  information in exchange for commercial incentives: An application of an
  extended theory of planned behavior}.
\newblock \bibinfo{journal}{\emph{Cyberpsychology, Behavior, and Social
  Networking}} \bibinfo{volume}{16}, \bibinfo{number}{2}
  (\bibinfo{year}{2013}), \bibinfo{pages}{81--87}.
\newblock


\bibitem[\protect\citeauthoryear{John, Acquisti, and Loewenstein}{John
  et~al\mbox{.}}{2011}]%
        {john2011strangers}
\bibfield{author}{\bibinfo{person}{Leslie~K John}, \bibinfo{person}{Alessandro
  Acquisti}, {and} \bibinfo{person}{George Loewenstein}.}
  \bibinfo{year}{2011}\natexlab{}.
\newblock \showarticletitle{Strangers on a plane: Context-dependent willingness
  to divulge sensitive information}.
\newblock \bibinfo{journal}{\emph{Journal of consumer research}}
  \bibinfo{volume}{37}, \bibinfo{number}{5} (\bibinfo{year}{2011}),
  \bibinfo{pages}{858--873}.
\newblock


\bibitem[\protect\citeauthoryear{Joshaghani, Black, Sherman, and
  Mehrpouyan}{Joshaghani et~al\mbox{.}}{2019}]%
        {joshaghani2019formal}
\bibfield{author}{\bibinfo{person}{Rezvan Joshaghani}, \bibinfo{person}{Stacy
  Black}, \bibinfo{person}{Elena Sherman}, {and} \bibinfo{person}{Hoda
  Mehrpouyan}.} \bibinfo{year}{2019}\natexlab{}.
\newblock \showarticletitle{Formal specification and verification of
  user-centric privacy policies for ubiquitous systems}. In
  \bibinfo{booktitle}{\emph{Proceedings of the 23rd International Database
  Applications \& Engineering Symposium}}. \bibinfo{pages}{1--10}.
\newblock


\bibitem[\protect\citeauthoryear{Knijnenburg and Kobsa}{Knijnenburg and
  Kobsa}{2014}]%
        {knijnenburg2014increasing}
\bibfield{author}{\bibinfo{person}{Bart~Piet Knijnenburg} {and}
  \bibinfo{person}{Alfred Kobsa}.} \bibinfo{year}{2014}\natexlab{}.
\newblock \showarticletitle{Increasing sharing tendency without reducing
  satisfaction: finding the best privacy-settings user interface for social
  networks}.
\newblock  (\bibinfo{year}{2014}).
\newblock


\bibitem[\protect\citeauthoryear{Kong, Gao, Chen, and Clarke}{Kong
  et~al\mbox{.}}{2015}]%
        {kong2015dreach}
\bibfield{author}{\bibinfo{person}{Soonho Kong}, \bibinfo{person}{Sicun Gao},
  \bibinfo{person}{Wei Chen}, {and} \bibinfo{person}{Edmund Clarke}.}
  \bibinfo{year}{2015}\natexlab{}.
\newblock \showarticletitle{dReach: $\delta$-reachability analysis for hybrid
  systems}. In \bibinfo{booktitle}{\emph{International Conference on TOOLS and
  Algorithms for the Construction and Analysis of Systems}}. Springer,
  \bibinfo{pages}{200--205}.
\newblock


\bibitem[\protect\citeauthoryear{Krishnan and Vorobyov}{Krishnan and
  Vorobyov}{2013}]%
        {krishnan2013enforcement}
\bibfield{author}{\bibinfo{person}{Padmanabhan Krishnan} {and}
  \bibinfo{person}{Kostyantyn Vorobyov}.} \bibinfo{year}{2013}\natexlab{}.
\newblock \showarticletitle{Enforcement of privacy requirements}. In
  \bibinfo{booktitle}{\emph{IFIP International Information Security
  Conference}}. Springer, \bibinfo{pages}{272--285}.
\newblock


\bibitem[\protect\citeauthoryear{Kurkovsky, Rivera, and Bhalodi}{Kurkovsky
  et~al\mbox{.}}{2007}]%
        {kurkovsky2007classification}
\bibfield{author}{\bibinfo{person}{O~Rivera Kurkovsky}, \bibinfo{person}{Oscar
  Rivera}, {and} \bibinfo{person}{Jay Bhalodi}.}
  \bibinfo{year}{2007}\natexlab{}.
\newblock \showarticletitle{Classification of privacy management techniques in
  pervasive computing}.
\newblock \bibinfo{journal}{\emph{International Journal of u-and e-Service,
  Science and Technology}} \bibinfo{volume}{11}, \bibinfo{number}{1}
  (\bibinfo{year}{2007}), \bibinfo{pages}{55--71}.
\newblock


\bibitem[\protect\citeauthoryear{Larsen, Pettersson, and Yi}{Larsen
  et~al\mbox{.}}{1997}]%
        {larsen1997uppaal}
\bibfield{author}{\bibinfo{person}{Kim~G Larsen}, \bibinfo{person}{Paul
  Pettersson}, {and} \bibinfo{person}{Wang Yi}.}
  \bibinfo{year}{1997}\natexlab{}.
\newblock \showarticletitle{UPPAAL in a nutshell}.
\newblock \bibinfo{journal}{\emph{International journal on software tools for
  technology transfer}} \bibinfo{volume}{1}, \bibinfo{number}{1-2}
  (\bibinfo{year}{1997}), \bibinfo{pages}{134--152}.
\newblock


\bibitem[\protect\citeauthoryear{Laufer and Wolfe}{Laufer and Wolfe}{1977}]%
        {laufer1977privacy}
\bibfield{author}{\bibinfo{person}{Robert~S Laufer} {and}
  \bibinfo{person}{Maxine Wolfe}.} \bibinfo{year}{1977}\natexlab{}.
\newblock \showarticletitle{Privacy as a concept and a social issue: A
  multidimensional developmental theory}.
\newblock \bibinfo{journal}{\emph{Journal of social Issues}}
  \bibinfo{volume}{33}, \bibinfo{number}{3} (\bibinfo{year}{1977}),
  \bibinfo{pages}{22--42}.
\newblock


\bibitem[\protect\citeauthoryear{Lederer, Mankoff, and Dey}{Lederer
  et~al\mbox{.}}{2003}]%
        {lederer2003wants}
\bibfield{author}{\bibinfo{person}{Scott Lederer}, \bibinfo{person}{Jennifer
  Mankoff}, {and} \bibinfo{person}{Anind~K Dey}.}
  \bibinfo{year}{2003}\natexlab{}.
\newblock \showarticletitle{Who wants to know what when? privacy preference
  determinants in ubiquitous computing}. In \bibinfo{booktitle}{\emph{CHI'03
  extended abstracts on Human factors in computing systems}}.
  \bibinfo{pages}{724--725}.
\newblock


\bibitem[\protect\citeauthoryear{Lipford, Besmer, and Watson}{Lipford
  et~al\mbox{.}}{2008}]%
        {lipford2008understanding}
\bibfield{author}{\bibinfo{person}{Heather~Richter Lipford},
  \bibinfo{person}{Andrew Besmer}, {and} \bibinfo{person}{Jason Watson}.}
  \bibinfo{year}{2008}\natexlab{}.
\newblock \showarticletitle{Understanding Privacy Settings in Facebook with an
  Audience View.}
\newblock \bibinfo{journal}{\emph{UPSEC}}  \bibinfo{volume}{8}
  (\bibinfo{year}{2008}), \bibinfo{pages}{1--8}.
\newblock


\bibitem[\protect\citeauthoryear{Liu, Gummadi, Krishnamurthy, and Mislove}{Liu
  et~al\mbox{.}}{2011}]%
        {liu2011analyzing}
\bibfield{author}{\bibinfo{person}{Yabing Liu}, \bibinfo{person}{Krishna~P
  Gummadi}, \bibinfo{person}{Balachander Krishnamurthy}, {and}
  \bibinfo{person}{Alan Mislove}.} \bibinfo{year}{2011}\natexlab{}.
\newblock \showarticletitle{Analyzing facebook privacy settings: user
  expectations vs. reality}. In \bibinfo{booktitle}{\emph{Proceedings of the
  2011 ACM SIGCOMM conference on Internet measurement conference}}.
  \bibinfo{pages}{61--70}.
\newblock


\bibitem[\protect\citeauthoryear{Lu, Huang, and Ke}{Lu et~al\mbox{.}}{2014}]%
        {lu2014verification}
\bibfield{author}{\bibinfo{person}{Jiajun Lu}, \bibinfo{person}{Zhiqiu Huang},
  {and} \bibinfo{person}{Changbo Ke}.} \bibinfo{year}{2014}\natexlab{}.
\newblock \showarticletitle{Verification of Behavior-aware Privacy Requirements
  in Web Services Composition.}
\newblock \bibinfo{journal}{\emph{JSW}} \bibinfo{volume}{9},
  \bibinfo{number}{4} (\bibinfo{year}{2014}), \bibinfo{pages}{944--951}.
\newblock


\bibitem[\protect\citeauthoryear{Lwin and Williams}{Lwin and Williams}{2003}]%
        {lwin2003model}
\bibfield{author}{\bibinfo{person}{May~O Lwin} {and} \bibinfo{person}{Jerome~D
  Williams}.} \bibinfo{year}{2003}\natexlab{}.
\newblock \showarticletitle{A model integrating the multidimensional
  developmental theory of privacy and theory of planned behavior to examine
  fabrication of information online}.
\newblock \bibinfo{journal}{\emph{Marketing Letters}} \bibinfo{volume}{14},
  \bibinfo{number}{4} (\bibinfo{year}{2003}), \bibinfo{pages}{257--272}.
\newblock


\bibitem[\protect\citeauthoryear{Macovei}{Macovei}{2015}]%
        {macovei2015determinants}
\bibfield{author}{\bibinfo{person}{Octav-Ionu{\c{t}} Macovei}.}
  \bibinfo{year}{2015}\natexlab{}.
\newblock \showarticletitle{Determinants of consumers’ pro-environmental
  behavior--toward an integrated model}.
\newblock \bibinfo{journal}{\emph{Journal of Danubian Studies and Research}}
  \bibinfo{volume}{5}, \bibinfo{number}{2} (\bibinfo{year}{2015}).
\newblock


\bibitem[\protect\citeauthoryear{Mazzia, LeFevre, and Adar}{Mazzia
  et~al\mbox{.}}{2012}]%
        {mazzia2012pviz}
\bibfield{author}{\bibinfo{person}{Alessandra Mazzia}, \bibinfo{person}{Kristen
  LeFevre}, {and} \bibinfo{person}{Eytan Adar}.}
  \bibinfo{year}{2012}\natexlab{}.
\newblock \showarticletitle{The pviz comprehension tool for social network
  privacy settings}. In \bibinfo{booktitle}{\emph{Proceedings of the Eighth
  Symposium on Usable Privacy and Security}}. \bibinfo{pages}{1--12}.
\newblock


\bibitem[\protect\citeauthoryear{Mehdy, Ekstrand, Knijnenburg, and
  Mehrpouyan}{Mehdy et~al\mbox{.}}{2021}]%
        {mehdy2021privacy}
\bibfield{author}{\bibinfo{person}{AK Mehdy}, \bibinfo{person}{Michael~D
  Ekstrand}, \bibinfo{person}{Bart~P Knijnenburg}, {and} \bibinfo{person}{Hoda
  Mehrpouyan}.} \bibinfo{year}{2021}\natexlab{}.
\newblock \showarticletitle{Privacy as a Planned Behavior: Effects of
  Situational Factors on Privacy Perceptions and Plans}.
\newblock \bibinfo{journal}{\emph{UMAP’21, June 21--25, 2021, Utrecht,
  Netherlands{\copyright} 2021 Association for Computing Machinery.}}
  (\bibinfo{year}{2021}).
\newblock


\bibitem[\protect\citeauthoryear{Mehdy and Mehrpouyan}{Mehdy and
  Mehrpouyan}{2020}]%
        {mehdy2020user}
\bibfield{author}{\bibinfo{person}{AKM~Nuhil Mehdy} {and} \bibinfo{person}{Hoda
  Mehrpouyan}.} \bibinfo{year}{2020}\natexlab{}.
\newblock \showarticletitle{A User-Centric and Sentiment Aware
  Privacy-Disclosure Detection Framework based on Multi-input Neural Network.}.
  In \bibinfo{booktitle}{\emph{PrivateNLP@ WSDM}}. \bibinfo{pages}{21--26}.
\newblock


\bibitem[\protect\citeauthoryear{Mehdy, Kennington, and Mehrpouyan}{Mehdy
  et~al\mbox{.}}{2019}]%
        {nuhil2019privacy}
\bibfield{author}{\bibinfo{person}{Nuhil Mehdy}, \bibinfo{person}{Casey
  Kennington}, {and} \bibinfo{person}{Hoda Mehrpouyan}.}
  \bibinfo{year}{2019}\natexlab{}.
\newblock \showarticletitle{Privacy Disclosures Detection in Natural-Language
  Text Through Linguistically-motivated Artificial Neural Network}. In
  \bibinfo{booktitle}{\emph{2nd EAI International Conference on Security and
  Privacy in New Computing Environments}}. EAI.
\newblock


\bibitem[\protect\citeauthoryear{Mehrpouyan, Azpiazu, and Pera}{Mehrpouyan
  et~al\mbox{.}}{2017}]%
        {mehrpouyan2017measuring}
\bibfield{author}{\bibinfo{person}{Hoda Mehrpouyan},
  \bibinfo{person}{Ion~Madrazo Azpiazu}, {and} \bibinfo{person}{Maria~Soledad
  Pera}.} \bibinfo{year}{2017}\natexlab{}.
\newblock \showarticletitle{Measuring personality for automatic elicitation of
  privacy preferences}. In \bibinfo{booktitle}{\emph{2017 IEEE Symposium on
  Privacy-Aware Computing (PAC)}}. IEEE, \bibinfo{pages}{84--95}.
\newblock


\bibitem[\protect\citeauthoryear{Nissenbaum}{Nissenbaum}{2004}]%
        {nissenbaum2004privacy}
\bibfield{author}{\bibinfo{person}{Helen Nissenbaum}.}
  \bibinfo{year}{2004}\natexlab{}.
\newblock \showarticletitle{Privacy as contextual integrity}.
\newblock \bibinfo{journal}{\emph{Wash. L. Rev.}}  \bibinfo{volume}{79}
  (\bibinfo{year}{2004}), \bibinfo{pages}{119}.
\newblock


\bibitem[\protect\citeauthoryear{Noam}{Noam}{2015}]%
        {noam}
\bibfield{author}{\bibinfo{person}{Noam}.} \bibinfo{year}{2015}\natexlab{}.
\newblock \bibinfo{title}{{Noam is a JavaScript library for working with
  automata and formal grammars for regular and context-free languages}}.
\newblock \bibinfo{howpublished}{https://github.com/izuzak/noam}.
\newblock
\newblock
\shownote{[Online; accessed 10-May-2021].}


\bibitem[\protect\citeauthoryear{Osborn, Sandhu, and Munawer}{Osborn
  et~al\mbox{.}}{2000}]%
        {osborn2000configuring}
\bibfield{author}{\bibinfo{person}{Sylvia Osborn}, \bibinfo{person}{Ravi
  Sandhu}, {and} \bibinfo{person}{Qamar Munawer}.}
  \bibinfo{year}{2000}\natexlab{}.
\newblock \showarticletitle{Configuring role-based access control to enforce
  mandatory and discretionary access control policies}.
\newblock \bibinfo{journal}{\emph{ACM Transactions on Information and System
  Security (TISSEC)}} \bibinfo{volume}{3}, \bibinfo{number}{2}
  (\bibinfo{year}{2000}), \bibinfo{pages}{85--106}.
\newblock


\bibitem[\protect\citeauthoryear{Petronio}{Petronio}{2015}]%
        {petronio2015communication}
\bibfield{author}{\bibinfo{person}{Sandra Petronio}.}
  \bibinfo{year}{2015}\natexlab{}.
\newblock \showarticletitle{Communication privacy management theory}.
\newblock \bibinfo{journal}{\emph{The international encyclopedia of
  interpersonal communication}} (\bibinfo{year}{2015}), \bibinfo{pages}{1--9}.
\newblock


\bibitem[\protect\citeauthoryear{Shen and Hong}{Shen and Hong}{2006}]%
        {shen2006attribute}
\bibfield{author}{\bibinfo{person}{Hai-bo Shen} {and} \bibinfo{person}{Fan
  Hong}.} \bibinfo{year}{2006}\natexlab{}.
\newblock \showarticletitle{An attribute-based access control model for web
  services}. In \bibinfo{booktitle}{\emph{2006 Seventh International Conference
  on Parallel and Distributed Computing, Applications and Technologies
  (PDCAT'06)}}. IEEE, \bibinfo{pages}{74--79}.
\newblock


\bibitem[\protect\citeauthoryear{Simonson and Tversky}{Simonson and
  Tversky}{1992}]%
        {simonson1992choice}
\bibfield{author}{\bibinfo{person}{Itamar Simonson} {and} \bibinfo{person}{Amos
  Tversky}.} \bibinfo{year}{1992}\natexlab{}.
\newblock \showarticletitle{Choice in context: Tradeoff contrast and
  extremeness aversion}.
\newblock \bibinfo{journal}{\emph{Journal of marketing research}}
  \bibinfo{volume}{29}, \bibinfo{number}{3} (\bibinfo{year}{1992}),
  \bibinfo{pages}{281--295}.
\newblock


\bibitem[\protect\citeauthoryear{Tesfay, Hofmann, Nakamura, Kiyomoto, and
  Serna}{Tesfay et~al\mbox{.}}{2018}]%
        {tesfay2018read}
\bibfield{author}{\bibinfo{person}{Welderufael~B Tesfay},
  \bibinfo{person}{Peter Hofmann}, \bibinfo{person}{Toru Nakamura},
  \bibinfo{person}{Shinsaku Kiyomoto}, {and} \bibinfo{person}{Jetzabel Serna}.}
  \bibinfo{year}{2018}\natexlab{}.
\newblock \showarticletitle{I read but don't agree: Privacy policy benchmarking
  using machine learning and the eu gdpr}. In
  \bibinfo{booktitle}{\emph{Companion Proceedings of the The Web Conference
  2018}}. \bibinfo{pages}{163--166}.
\newblock


\bibitem[\protect\citeauthoryear{Van~Schaik}{Van~Schaik}{1999}]%
        {van1999involving}
\bibfield{author}{\bibinfo{person}{Paul Van~Schaik}.}
  \bibinfo{year}{1999}\natexlab{}.
\newblock \showarticletitle{Involving users in the specification of
  functionality using scenarios and model-based evaluation}.
\newblock \bibinfo{journal}{\emph{Behaviour \& Information Technology}}
  \bibinfo{volume}{18}, \bibinfo{number}{6} (\bibinfo{year}{1999}),
  \bibinfo{pages}{455--466}.
\newblock


\bibitem[\protect\citeauthoryear{West, Mayhorn, Hardee, and Mendel}{West
  et~al\mbox{.}}{2009}]%
        {west2009weakest}
\bibfield{author}{\bibinfo{person}{Ryan West}, \bibinfo{person}{Christopher
  Mayhorn}, \bibinfo{person}{Jefferson Hardee}, {and} \bibinfo{person}{Jeremy
  Mendel}.} \bibinfo{year}{2009}\natexlab{}.
\newblock \showarticletitle{The weakest link: A psychological perspective on
  why users make poor security decisions}.
\newblock In \bibinfo{booktitle}{\emph{Social and Human elements of information
  security: Emerging Trends and countermeasures}}. \bibinfo{publisher}{IGI
  Global}, \bibinfo{pages}{43--60}.
\newblock


\bibitem[\protect\citeauthoryear{Westin}{Westin}{1968}]%
        {westin1968privacy}
\bibfield{author}{\bibinfo{person}{Alan~F Westin}.}
  \bibinfo{year}{1968}\natexlab{}.
\newblock \showarticletitle{Privacy and freedom}.
\newblock \bibinfo{journal}{\emph{Washington and Lee Law Review}}
  \bibinfo{volume}{25}, \bibinfo{number}{1} (\bibinfo{year}{1968}),
  \bibinfo{pages}{166}.
\newblock


\bibitem[\protect\citeauthoryear{Xiao and Tao}{Xiao and Tao}{2006}]%
        {xiao2006personalized}
\bibfield{author}{\bibinfo{person}{Xiaokui Xiao} {and} \bibinfo{person}{Yufei
  Tao}.} \bibinfo{year}{2006}\natexlab{}.
\newblock \showarticletitle{Personalized privacy preservation}. In
  \bibinfo{booktitle}{\emph{Proceedings of the 2006 ACM SIGMOD international
  conference on Management of data}}. \bibinfo{pages}{229--240}.
\newblock


\bibitem[\protect\citeauthoryear{Yao and Linz}{Yao and Linz}{2008}]%
        {yao2008predicting}
\bibfield{author}{\bibinfo{person}{Mike~Z Yao} {and} \bibinfo{person}{Daniel~G
  Linz}.} \bibinfo{year}{2008}\natexlab{}.
\newblock \showarticletitle{Predicting self-protections of online privacy}.
\newblock \bibinfo{journal}{\emph{CyberPsychology \& Behavior}}
  \bibinfo{volume}{11}, \bibinfo{number}{5} (\bibinfo{year}{2008}),
  \bibinfo{pages}{615--617}.
\newblock


\bibitem[\protect\citeauthoryear{Yoo, Jee, and Cha}{Yoo et~al\mbox{.}}{2009}]%
        {yoo2009formal}
\bibfield{author}{\bibinfo{person}{Junbeom Yoo}, \bibinfo{person}{Eunkyoung
  Jee}, {and} \bibinfo{person}{Sungdeok Cha}.} \bibinfo{year}{2009}\natexlab{}.
\newblock \showarticletitle{Formal modeling and verification of safety-critical
  software}.
\newblock \bibinfo{journal}{\emph{IEEE software}} \bibinfo{volume}{26},
  \bibinfo{number}{3} (\bibinfo{year}{2009}), \bibinfo{pages}{42--49}.
\newblock


\end{thebibliography}

\appendix

\section{Survey Interface 1}
\label{appendix:survey-ui}
\begin{figure}
    \includegraphics[width=\linewidth]{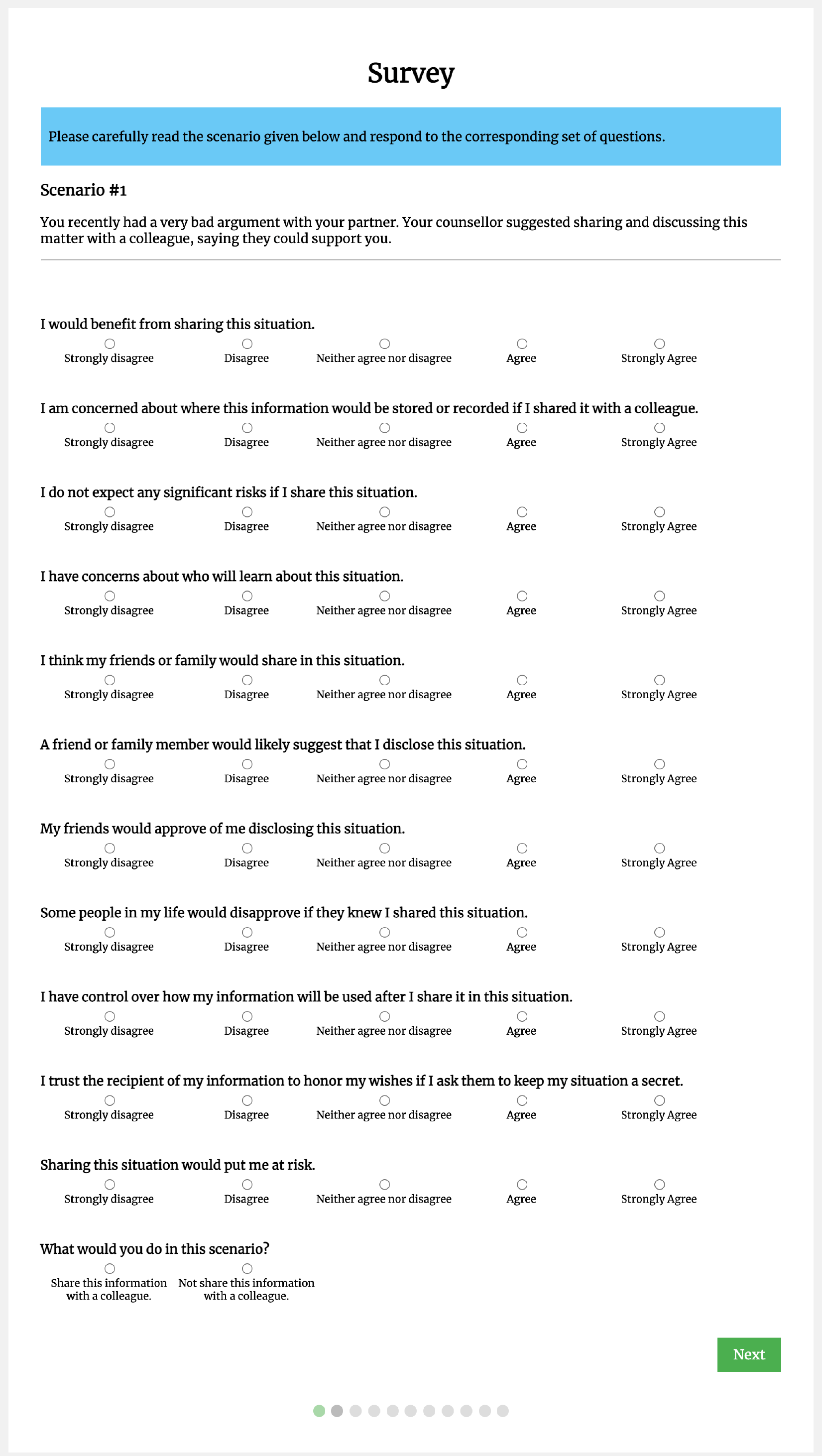}
    \caption{Screenshot of the Survey System Representing 1 of 8 Random Scenarios Given to a Participant.}
    \label{figure:app-sui}
\end{figure}

Screenshot of the survey system representing 1 of 8 random scenarios given to a participant (Figure \ref{figure:app-sui}).

\newpage
\section{Survey Interface 2}
\label{appendix:survey-ui-ga}
\begin{figure}
    \includegraphics[width=\linewidth]{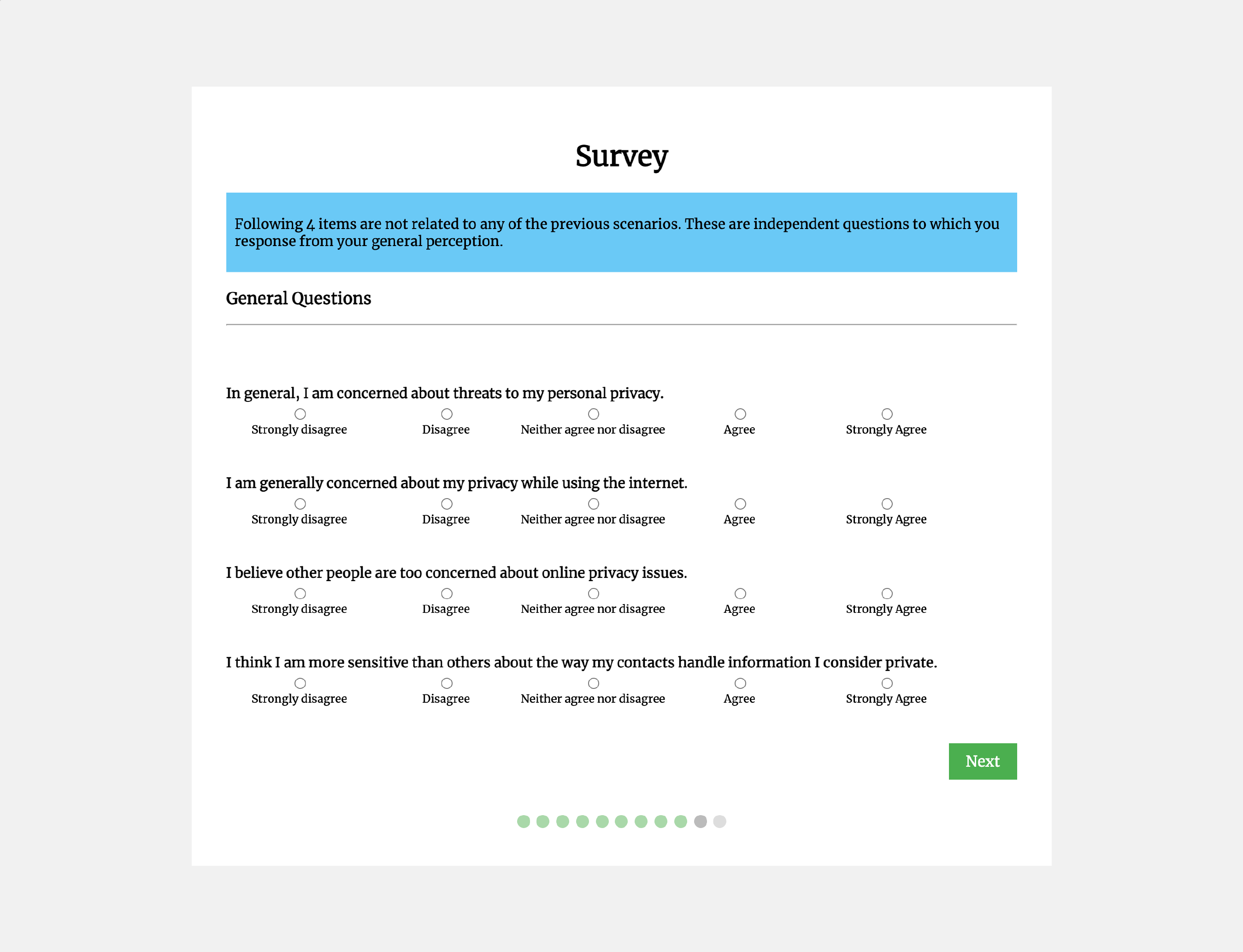}
    \caption{Screenshot of the survey system representing the general attitude questions given to a participant at the end of the survey.}
    \label{figure:app-sui-ga}
\end{figure}

Screenshot of the Survey System Representing the General Attitude Questions Given to a Participant at the end of the Survey (Figure \ref{figure:app-sui-ga}).

\end{document}